\documentclass[sts,preprint]{imsart}

\RequirePackage{amsthm,amsmath,amsfonts,amssymb}
\RequirePackage[numbers]{natbib}
\usepackage{graphicx}
\usepackage{caption}
\usepackage{subcaption}
\usepackage{hhline}

\startlocaldefs

\makeatletter
\renewcommand{\paragraph}{\@startsection{paragraph}{4}{0ex}%
   {-3.25ex plus -1ex minus -0.2ex}%
   {1.5ex plus 0.2ex}%
   {\normalfont\normalsize\bfseries}}
\makeatother

\stepcounter{secnumdepth}
\stepcounter{tocdepth}

\endlocaldefs

\begin{document}

\begin{frontmatter}
\title{Protocols for Observational Studies: Methods and Open Problems\thanks{This paper is based on my IMS Medallion Lecture on August 9th, 2022 at the 2022 Joint Statistical Meetings in Washington D.C.}}

\runtitle{Protocols for Observational Studies}

\begin{aug}
\author{Dylan S. Small \ead[label=e1]{dsmall@wharton.upenn.edu}}
\address{Dylan Small is Universal Furniture Professor, Department of Statistics and Data Science, The Wharton School, University of Pennsylvania, Philadelphia, PA, USA \printead{e1}.}

\end{aug}

\begin{abstract}

For learning about the causal effect of a treatment, a randomized controlled trial (RCT) is considered the gold standard.  However, randomizing treatment is sometimes unethical or infeasible, and instead an observational study may be conducted.  While some aspects of a well designed RCT cannot be replicated in an observational study, one aspect that can is to have a protocol with prespecified hypotheses about prespecified outcomes and a prespecified analysis.  We illustrate the value of protocols for observational studies in three applications {\textendash} the effect of playing high school football on later life mental functioning, the effect of police seizing a gun when arresting a domestic violence suspect on future domestic violence and the effect of mountaintop mining on health.  We then discuss methodologies for observational study protocols.  We discuss considerations for protocols that are similar between observational studies and RCTs, and considerations that are different.  The considerations that are different include (i) whether the protocol should be specified before treatment assignment is known or after; (ii) how multiple outcomes should be incorporated into the planned analysis and (iii) how subgroups should be incorporated into the planned analysis.  We conclude with discussion of a few open problems in the methodology of observational study protocols.
\end{abstract}


\end{frontmatter}

\section{Introduction}

For learning about the effect of taking a treatment, randomized controlled trials (RCTs) are considered the gold standard \cite{hill1979periodic,lawrence1989us}.  However, RCTs are often too costly or unethical, and observational studies are used instead.  In his seminal paper on observational studies, William Cochran suggested that when planning observational studies, look to RCTs as a guide \cite{cochran1965planning}.\footnote{See \cite{rubin2007design,hernan2016using} for recent expressions of this view}  The strength of a properly designed RCT comes from three sources according to Lincoln Moses \cite{moses1995measuring}:
\begin{enumerate}
\item {\it{Randomization}}: By randomly assigning people (or other units) to treatment $A$ or $B$, it creates a fair comparison between $A$ and $B$ in terms of pre-treatment covariates, e.g., the people given $A$ aren't systematically healthier or wealthier.
\item {\it{Identical Processes}}: It takes pains to apply to people treated by $A$ or $B$ all other processes in the same way, e.g., adjuvant therapies, follow-up, record keeping and outcome assessment.  Double blinding (blinding participants to what treatments they are receiving by using a placebo and blinding researchers to what treatments participants are receiving) helps to ensure identical processes.
\item {\it{Protocol}}: The whole undertaking is driven by a strong protocol with prespecified hypotheses about prespecified outcomes and a prespecified analysis.
\end{enumerate}
The first source of strength, randomization, is unique to an RCT.
The second source of strength, identical processes, is sometimes difficult to achieve in an observational study as investigators lack control over the environment.  For example, placebos are typically not possible in an observational study.  But the third source of strength, a protocol, has nothing to do with randomization or control over the environment, and can be made part of a well designed observational study.  This article will discuss methods and open problems for protocols in observational studies with special attention to aspects of protocols which differ for observational studies compared to randomized trials.

The value of having prespecified hypotheses about prespecified outcomes, a key aspect of a protocol, was understood by two pioneering investigators {\textendash} Daniel in his test of his vegetarian diet compared with meat in the bible's {\it{Book of Daniel}} and Lind in his 1747 test of treatments for scurvy \cite{wittes2018}.  For example, Daniel's hypothesis was that a 10-day diet of vegetarian food would not make a youth's face look worse (a noninferiority hypothesis)\cite{daniel1953}:
\begin{quote}
Then let our appearance and the appearance of the youths who eat the king's rich food be observed by you [Melzar, Prince of Eunuchs]
\end{quote}
But many observational studies today do not follow a clear protocol \cite{rubin2007design}.

An example of how not following a clear protocol is problematic is \cite{gautret2020hydroxychloroquine}, which claimed to provide evidence that hydroxychloroquine was effective for treating COVID-19.  While a protocol for the study was registered in the EU Clinical Trials register in which the primary endpoints were listed as COVID-19 detection at days 1, 4, 7 and 14 after treatment, in the published paper, the primary outcome was listed as virus detection at day 6 and results were given for days 3, 4, 5 and 6.  Among the results presented, the $p$-value was largest for day 4, the only day which was a pre-specified primary outcome.  It is concerning that the primary outcome was changed from the protocol to the published paper \cite{rosendaal2020review} and ``it is concerning that Day 4 is the only pre-specified day that was ultimately presented, and that is the one with the largest p-value of the presented results.''\cite{faybrittain2022}

Even when there are prespecified hypotheses about prespecified outcomes in an observational study, the analysis is often not prespecified.  A typical approach is to fit a regression model of the outcome on the treatment and measured confounders, where the coefficient on the treatment estimates the treatment effect.
Donald Rubin \cite{rubin2007design} criticized how this is done in practice:
\begin{quote}
[In observational studies in epidemiology and social science, the outcome data] are used over and over and over again to fit various models, try different transformations, look at results discarding influential outliers etc.  ‘Oh, I should have used five indicator variables for age rather than age as continuous, because the p-values for treatment effects greatly improved!’ How many reported analyses that we see in journals are ‘designed’ a priori rather than are the results of repeated and unreported exploratory analyses?
\end{quote}
Rubin proposes instead to design observational studies without looking at the outcome data:
\begin{quote}
Observational studies can and should be designed to approximate randomized experiments as closely as possible. In particular, observational studies should be designed using only background information to create subgroups of similar treated and control units, where ‘similar’ here refers to their distributions of background variables [i.e., covariates]. Of great importance, this activity should be conducted without any access to any outcome data, thereby assuring the objectivity of the design...Of course, objectivity is not the same as finding truth, but I believe that it is generally a necessary ingredient if we are to find
truth.
\end{quote}

The protocol of an observational study should describe the study's design and the plan for analysis.  In Section \ref{sec: design}, we provide a brief review of methods for the design and analysis of observational studies.  In Section \ref{sec: case studies}, we present three case studies from our own work of using protocols in observational studies, highlighting the value of the protocol in each case.  In Section \ref{sec: common features}, we discuss common features of good observational study and RCT protocols, and in Section \ref{sec: different.considerations}, we discuss differences.  In Section \ref{open.problems}, we discuss open problems for observational study protocols.

\section{Review: Design and Analysis of Observational Studies}
\label{sec: design}

The starting point for designing an observational study is to formulate the research question \cite{rosenbaum1999choice,cox2016design}.  
Next, the study population must be chosen and the treatment and control groups within this population defined.
The outcome(s) must also be chosen.   If more than one primary outcome will be considered, multiple testing should be controlled for \cite{tukey1991philosophy}.  The start and end time of follow-up for the outcomes may need to be considered.  A useful approach to designing observational studies can be to design them so they explicitly emulate a hypothetical randomized trial that would answer the question at hand \cite{hernan2016using}.  This helps avoid biases such the start of follow-up being different for the treatment and control groups \cite{hernan2016specifying}.

After choosing the outcomes, the investigator needs to choose what covariates will be adjusted for.  \cite{rosenbaum2002obsstudies} (\S 3.1.3), \cite{witte2019covariate}, \cite{vanderweele2019principles}, \cite{loh2021confounder} and \cite{guo2023confounder} discuss some approaches and considerations.  

Biases between the treatment and control groups in covariates that will be adjusted for are overt biases that we hope to remove by adjustment methods.  Such adjustment methods include regression, stratification, matching, inverse probability weighting and doubly robust estimation.\footnote{Propensity scores \citep{rosenbaum1983central} can play a useful role in all of these methods \cite{vansteelandt2014regression,rosenbaum2023propensity}.}
    See \cite{rosenbaum2002obsstudies,imbens2015causal,hernan2020causal} for textbook discussions. We will focus in this paper on the matching approach in which matched sets of treated and control units are constructed so that the measured covariates of the treated and control units within each set are similar, or if this is not possible, the covariates are counterbalanced across different sets \cite{stuart2010matching,hansen2011propensity,rosenbaum2020modern}.  As long as there are no unmeasured confounders (meaning that conditional on the covariates which have been adjusted for, the treatment is effectively randomly assigned), a fair comparison between treatment and control can be made by aggregating comparisons of treated and control within matched sets.  Matching coordinates well with specifying a protocol in that a simple and robust estimator of the treatment effect can be prespecified, e.g., (i) the average across matched sets of the within matched set difference between the averages of the treated and control outcomes or (ii) a linear regression of the outcome on the treatment, the covariates and the matched set indicators where the coefficient on the treatment is the estimated treatment effect.  Rubin \cite{rubin1979using} found in a simulation study that both estimators were robust for estimating the treatment effect even when the covariates had a nonlinear association with the outcome, with estimator (ii) being more efficient.  In contrast, a linear regression of the outcome on the treatment and covariates without matching was sometimes substantially biased when the covariates had a nonlinear association with the outcome.

Other adjustment methods besides matching can also coordinate well with a protocol.  Like matching, stratification and inverse probability weighting can be designed prior to looking at the outcome data and a robust estimator can be prespecified \cite{rosenbaum1984reducing,gruber2015ensemble,smith2022emulation}.  A regression approach that can coordinate well with a protocol is to use a machine learning type method (including Bayesian tree regression methods) that is designed to be black box, requiring no input from the investigator \cite{hill2011bayesian,van2011targeted,wager2018estimation,hahn2020bayesian}.
Doubly robust estimation, which combines a model for how the covariates are associated with the treatment and a model for how the covariates are associated with the outcome, can also be prespecified in a robust way, e.g., the treatment model can be estimated before looking at the outcome data and a black box machine learning method for the outcome model can be prespecified \cite{chernozhukov2018double}.  Perhaps the most challenging adjustment method to coordinate with a protocol is a traditional regression adjustment that involves exploring the outcome data to choose a regression model.  But one approach is to prespecify as much as possible how the data will be explored and a method for how inferences will account for the data exploration \cite{kuchibhotla2022post}.  Although in this paper we focus on observational study protocols using matching, we believe that many of the issues are also relevant for observational study protocols using other adjustment approaches (for example, see Section 5 of the Supplemental Materials).



Although overt biases in an observational study can often be well controlled for by adjustment methods, there is typically concern about hidden biases from unmeasured confounders (unmeasured covariates that are related to both the treatment and the outcome conditional on the measured covariates, see \cite{vanderweele2013definition} for a formal definition).  For example, consider Hammond's \cite{hammond1964smoking} study of smoking and lung cancer which matched 36,975 heavy smokers to nonsmokers on sociodemographic and health characteristics.  Sir R.A. Fisher raised the concern that there might be hidden bias from an unmeasured genetic variant that predisposes a person to smoke and to get lung cancer.  However, claims about hidden bias do not become credible merely because the covariates involved were not observed \cite{rosenbaum2002obsstudies}.  The issue should be explored through a sensitivity analysis that asks what magnitude of unmeasured confounding would need to be present to materially alter the study's causal conclusions.  Many sensitivity analysis methods that go along with different adjustment methods have been developed including \cite{hosman2010sensitivity,ding2016sensitivity,cinelli2020making,zhang2020calibrated,bonvini2022sensitivity}; see \cite{fogarty2023sensitivity} for a review.  A sensitivity analysis that goes along with matching is Rosenbaum's $\Gamma$ sensitivity analysis method \cite{rosenbaum2002obsstudies} which is indexed by a parameter $\Gamma >1$ that says two subjects matched for observed covariates may differ in their odds of exposure to treatment by at most a factor of $\Gamma$.  When there is no unmeasured confounding such as in a randomized trial, $\Gamma =1$.  When there is an unmeasured confounder that might double the odds of treatment, $\Gamma =2$.  When there is an unmeasured confounder that might triple the odds of treatment, $\Gamma =3$.  At each value of $\Gamma$, there is a range of possible values for an inference quantity, say a $p$-value, and the maximum can be computed \cite{rosenbaum2002obsstudies}.  For example, in Hammond's \cite{hammond1964smoking} study of smoking and lung cancer, there were 122 pairs in which exactly one person died of lung cancer (in the three years of followup) and in 110 of these, the person dying was a smoker.  The $p$-value for the usual McNemar's test (i.e, McNemar's test with $\Gamma =1$) is less than $0.0001$ so if there was no unmeasured confounding, this would be strong evidence that smoking causes lung cancer.  For $\Gamma =2$ or $3$, the maximum $p$-value is still less than $0.0001$; for $\Gamma =4$, it is $0.0036$; for $\Gamma =5$, it is $0.03$ and for $\Gamma =6$, it is $0.10$.  Thus, even if there were unmeasured confounding that quintupled the odds of smoking ($\Gamma =5$), there would still be evidence that smoking causes lung cancer (maximum $p=0.03$) although if was unmeasured confounding that sextupled the odds of smoking ($\Gamma =6$), there might no longer be evidence (maximum $p=0.10$) \footnote{The sensitivity analysis can be computed in \texttt{R} using the \texttt{binarysens} function in the \texttt{rbounds} package}.  The finding that smoking causes lung cancer is insensitive to a fairly large amount of unmeasured confounding.  

The maximum $p$-value for a given $\Gamma$ assumes the unmeasured confounding is maximally associated with the outcome.  We can instead allow for the unmeasured confounding to have limited associations with both the treatment and the outcome, a parameter $\Lambda$ controlling the relationship between exposure to the treatment (say smoking) and an unobserved covariate (say a genetic variant) and a parameter $\Delta$ controlling the relationship between the outcome the subject would experience if given the control (say whether a person would get lung cancer if she did not smoke) and the same unobserved covariate.  In matched pairs, the maximum $p$-value for a given ($\Lambda$,$\Delta$) pair corresponds to the maximum $p$-value for the following $\Gamma$, $
\Gamma = (\Lambda\Delta +1)/(\Lambda + \Delta )$.  Thus, one only needs to compute the $\Gamma$ sensitivity analysis and one automatically learns about the $(\Lambda ,\Delta )$ sensitivity analysis \cite{rosenbaum2009amplification}.  For example, an unobserved covariate that increases the odds of smoking 9.9 times and the odds of dying from lung cancer 9.9 times would correspond with $\Gamma =5$, and so have a maximum $p$-value of 0.03.


A sensitivity analysis can show that only large hidden biases would invalidate an observational study's conclusions but cannot rule out such large biases.  Large biases can be made less likely by careful choice of the circumstances of an observational study \citep{rosenbaum1999choice}.  For example, \cite{lehman1987long} studied the effect of the loss of a spouse or child in a car crash on long term mental health \cite{lehman1987long}
The study compared drivers who lost a spouse or child in a crash that was not the driver's fault to matched controls who came to renew their drivers' license.  Rosenbaum says \cite{rosenbaum1999choice},
\begin{quote} the restriction of attention to drivers who were not at fault reduced somewhat the magnitude of plausible hidden biases, since, for example, alcohol abuse is related to accidents in which one is at fault and also related to psychiatric outcomes.
\end{quote}

The threat from hidden biases to a study's validity can also be reduced by anticipating potential hidden biases and actively collecting data that would reveal those biases if present \cite{campbell1957factors}.  Examples of such data, called quasi-experimental devices, are multiple control groups that vary in their level of a worrisome unmeasured covariate and negative control outcomes that are thought to be unaffected by the treatment but highly associated with a worrisome unmeasured covariate \cite{cook2002experimental,rosenbaum2002obsstudies}.

We have provided a brief review of aspects of the design and analysis of an observational study, see \cite{rosenbaum1999choice,cook2002experimental,rosenbaum2010design,imbens2015causal,cox2016design,hernan2020causal} for more detailed discussion.  Whatever the design and planned analysis, we argue that it is beneficial to describe it in a written protocol before starting the analysis.  Ideally, the protocol should be made public and circulated for critical comment.  Rosenbaum \cite{rosenbaum2002obsstudies} writes:  
\begin{quote}
A small oversight, easily corrected in the planning stage, may be an insurmountable problem at a later stage...Observational studies would, I believe, benefit from a written protocol and critical commentary.
\end{quote}

\section{Case Studies of Observational Study Protocols}
\label{sec: case studies}

\subsection{High School Football and Later Life Mental Functioning}

In \cite{deshpande2017association}, my collaborators and I studied the effect of playing American football on later life mental functioning (cognition and mental health).  The idea for the study came to me after watching the movie {\it{Concussion}}.  
The movie's hero is Dr. Bennet Omalu, a forensic pathologist who fights against the National Football League trying to suppress his research on brain damage suffered by professional football players from frequent hits to the head while playing.  I became curious about this topic and found that most research was on professional football players.  While professional football is played by a select few, high school football is played by about a million boys a year.  I realized that a data set I had previously worked with, the Wisconsin Longitudinal Study (WLS) \cite{herd2014cohort}, contained information on high school activities such as playing football and later life mental functioning.  I dreamed of making an important public health finding as a statistician like some of my statistical heroes William Cochran, Jerome Cornfield and Austin Bradford Hill did on the harmful effects of smoking, and maybe getting mentioned in my favorite newspaper, the {\it{New York Times}}.  I talked with two Ph.D. students in my department who I knew were sports fans and interested in causal inference, Sameer Deshpande and Raiden Hasegawa, and they were interested and led the study.  We recruited experts on adolescent sports, brain injury, concussions, geriatrics, neurology and  observational studies  {\textendash} Christina Master, Amanda Rabinowitz, John Whyte, Jason Karlawish, Carol Roan and Mike Baiocchi {\textendash} to collaborate with us as well as a high school student interested in research, Andrew Tabatabaei.

\subsubsection{Protocol}

We discussed with our collaborators what should be the primary and secondary outcomes.  The WLS administered a battery of cognitive tests and mental health surveys in 1993, between 2003 and 2005, and between 2011 and 2013, when the subjects were approximately 53, 65, and 72 years old, respectively.  In consultation with our collaborators, we decided on two primary outcomes, depression at age 65 and cognition at age 65.  To measure depression, we decided to use the Center for Epidemiologic Studies Depression (CES-D) scale \cite{radloff1977ces}.  For cognition, we averaged the z-scores for a measure of memory/attention (delayed word recall\footnote{While answering a phone survey, the respondent is told ``Part of this study is concerned with people's memory.  I'll read a set of 10 words and ask you to recall as many as you can.  Please listen carefully as I read the set of words. I'm not allowed to repeat any of the words, so it's important that you can hear me very well.  When I finish, I will ask you to recall aloud as many of the words as you can, in any order.  Is this clear?  Ok.  The list is...".
[ten words are read, e.g., hotel, river, skin, tree, gold, market, paper, child, king, and book.]
Once the interviewer finishes, the respondent repeats back as many words as he or she can remember and the interviewer records the repeated words. Then, about twelve minutes later, the respondent is told, ``A little while ago, I read you a list of words and you repeated the ones you could remember.  Please tell me any of the words that you remember now.''  The number of remembered words is the delayed word recall score.}) and a measure of verbal fluency (letter fluency\footnote{The respondent is asked to think of as many words as they can beginning with a certain letter in 60 seconds.  The letter is randomly chosen to be either ``L'' or ``F.''}).  These two cognitive domains for measuring cognition were chosen because they are most consistent with the National Institute of Health (NIH) recommendations for Common Data Elements in
Traumatic Brain Injury research \cite{wilde2010recommendations} and have demonstrated the greatest sensitivity to sports-related concussion \cite{belanger2005neuropsychological}.
In addition to the primary outcomes, we decided to consider several secondary outcomes such as anger and anxiety \cite{deshpande2017association}.

Our collaborators also helped us identify potential confounding variables available in the WLS that are shown in Table \ref{balance.table.football}.  Several of these variables differed between football players and non-football players such as planned future education (football players $2.34$ years vs. non-football players $1.79$ years) and participation in school government (football players 36\% vs. non-football players 15\%).  To control for the potential confounding variables, we matched football players to non-football players (we only considered men since there were no women football players at this time) on these variables using full matching \cite{rosenbaum1991characterization,hansen2004full}.  The balance between the football players and matched controls on the matching variables is shown in Table \ref{balance.table.football}.  We reviewed the table with our collaborators and they felt the balance on the matched variables was adequate.  The balance between the football players and non-football players on IQ, planned education, family wealth, socioeconomic status of father's job and socioeconomic status of job aspired to were thought to be particularly important because of these variables' potential effects on cognition tests or depression.  After matching, the standardized differences for these variables were all less than $0.1$ which is considered desirable balance \cite{silber2013characteristics}.\footnote{The standardized difference is the average  difference in means between the treated (football players) and controls (non-football players) within matched sets (weighted according to the number of football players in each matched set) in terms of pooled within-group (treatment group and control group before matching) standard deviation units \cite{rosenbaum1985constructing}.}

\begin{table}[ht]
\centering
\caption{Covariate balance, before and after matching, in the football study.  Because matched sets vary in size,
means are weighted to reflect frequencies in the treated (football) group.  The standardized difference is the difference
in means divided by the standard deviation prior to matching \cite{rosenbaum1985constructing}.  Standardized differences above
0.2 are in \textbf{bold}.}
\begin{tabular}{|l|rrr|rr|}
 \hline
 & \multicolumn{3}{|c}{Weighted Mean} & \multicolumn{2}{|c|}{Standardized Difference} \\ \hline
& Treated & Matched  & All  & Before & After  \\
& & control & controls & matching & matching \\
 \hline
Covariate & & & &  \\ \hline
Duncan socioeconomic index of job & 577.67 & 541.94 & 560.13 & 0.16 & 0.08 \\
\ participant aspired to when high school senior & & & & & \\
High school size & 134.86 & 188.12 & 139.01 & \textbf{--0.43} & 0.03 \\
High school rank (quantile) & 48.81 & 45.12 & 47.45 & 0.14 & 0.05 \\
Parental income in 1957 (\$ 100) & 6938 & 6253 & 6522 & 0.10 & 0.06 \\
Parental education level & & & & & \\
\ Father & 9.91 & 9.76 & 9.84 & 0.04 & 0.02 \\
\ Mother & 10.80 & 10.50 & 10.72 & 0.11 & 0.03 \\
Duncan socioeconomic index of father's job & 355.30 & 339.87 & 348.06 & 0.07 & 0.03 \\
Planned future education (years) & 2.34 & 1.79 & 2.17 & \textbf{0.26} & 0.08 \\
IQ & 102.39 & 101.60 & 101.88 & 0.05 & 0.03 \\
Considered `outstanding student' by teacher (\%) & 12 & 10 & 12 & 0.04 & -0.01 \\
Participated in band, orchestra, & 34 & 29 & 33 & 0.11 & 0.02 \\
\ chorus or musical ensemble & & & & & \\
Participated in drama, speech or debate (\%) & 29 & 20 & 28 & \textbf{0.21} & 0.02 \\
Participated in school government (\%) & 36 & 15 & 30 & \textbf{0.49} & 0.14 \\
Participated in school publications (\%) & 23 & 15 & 23 & 0.18 & -0.01 \\
Father was a farmer (\%) & 18 & 19 & 18 & -0.01 & 0.00 \\
Planned to serve in the military (\%) & 27 & 27 & 27 & -0.01 & 0.00 \\
Attended Catholic high school (\%) & 7 & 11 & 8 & -0.17 & -0.05 \\
Lived with both parens (\%) & 91 & 91 & 90 & 0.02 & 0.06 \\
Mother working in 1957 & 37 & 36 & 3 & 0.02 & 0.03 \\
Teachers encouraged college & 58 & 45 & 5 & \textbf{0.25} & 0.05 \\
Parents encouraged college & 65 & 60 & 65 & 0.10 & -0.01 \\
Had friends who planned to go to college & 43 & 36 & 42 & 0.14 & 0.01 \\
Discussed future plans with teachers (\%) & & & & & \\
\ Not at all & 24 & 31 & -0.12 & -0.04 \\
\ Sometimes & 66 & 61 & 64 & 0.08 & 0.03 \\
\ Very much & 10 & 8 & 10 & 0.04 & 0.01 \\
Discussed future plans with teachers (\%) & & & & & \\
\ Not at all & 2 & 2 & 2 & -0.01 & 0.01 \\
\ Sometimes & 42 & 46 & 44 & -0.07 & -0.03 \\
\ Very much & 56 & 52 & 54 & 0.07 & 0.03 \\
Family wealth relative to community (\%) & & & & & \\
\ Considerably below average & 1 & 1 & 0 & 0.0 & 0.00 \\
\ Somewhat below average & 5 & 7 & 7 & -0.03 & -0.04 \\
\ Average & 70 & 70 & 68 & 0.00 & 0.04 \\
\ Somewhat above average & 22 & 21 & 22 & 0.02 & 0.00 \\
\ Considerably above average & 3 & 2 & 2 & 0.02 & 0.00 \\
Financial support from parents for college (\%) & & & & & \\
\ Cannot support & 28 & 33 & 31 & -0.07 & -0.05 \\
\ Can support, with some sacrifice & 58 & 54 & 6 & 0.06 & 0.04 \\
\ Easily support & 14 & 13 & 13 & 0.02 & 0.01 \\ \hline
\end{tabular}
\label{balance.table.football}
\end{table}%

Although our collaborators felt the balance on the matched variables created a fair comparison between the treated and matched control groups, looking at Table \ref{balance.table.football} sparked concern about two unmatched variables.  One collaborator expressed concern that the APOE $\epsilon$4 genetic variant, a known risk factor for dementia, was not matched on.  We examined the APOE $\epsilon$4 genetic variant and found it was balanced between the football players and matched controls. Another collaborator said, ``Depressed poets don't play football.''  People who choose to play high school sports may have personalities that make them less prone to depression (and would make them less prone even if they hadn't played sports).  To address this concern, we designed multiple control groups that systematically vary the unmeasured confounder of concern, interest in playing high school sports.  We considered two control groups, (i) men who played only a non-collision sport (e.g., baseball, basketball or track) and (ii) men who did not play a sport all.  If these control groups have similar outcomes, this provides evidence that personality associated with interest in playing sports is not an important confounder \cite{campbell1969prospective,rosenbaum1987role}.

Before looking at the outcome data, we prespecified an analysis plan.  To estimate the effect of the treatment (playing football), we planned to
regress each primary outcome on matched set indicators, covariates, and a treatment indicator.  To control for the multiple testing involved in having two primary outcomes, we prespecified that we would use Bonferroni and test each outcome at level $0.025$.
To consider multiple control groups while controlling the familywise error rate at $0.05$, we prespecified that we would use testing in order to control the familywise error rate at $0.025$ for each of the primary outcomes \cite{rosenbaum2008testing}.
While we prespecified that we would control the familywise error rate for the primary outcomes at $0.05$, we were not as strict for the secondary outcomes and instead planned to report marginal $p$-values and also whether the $p$-value is significant when controlling the false discovery rate at $0.05$ using the Benjamini-Hochberg procedure \cite{benjamini1995controlling}.  For all of the outcomes, we planned to do a sensitivity analysis for how sensitive the inferences are to unmeasured confounding using Rosenbaum's $\Gamma$ method \cite{rosenbaum2002obsstudies}.

Although we completed the protocol before using the outcomes to estimate any treatment effects, we did preexamine the outcomes in a way that was unlinked to the treatment to assess whether missingness of outcomes was a problem.  If football playing status affected the availability of outcomes (e.g. playing football increased the likelihood of dying young or early onset of debilitating cognitive impairment so that the subject
was unable to participate in the WLS surveys), any comparison of treated and control groups could be biased. However, we did not find evidence for such an effect.  Specifically, we fit logistic regression models to predict the availability of the primary outcomes given football playing status and all of our measured covariates in Table \ref{balance.table.football} and football playing status was not a significant predictor\footnote{This does not test for all sources of bias from missingness of outcomes.  For example, football players suffering early onset of debilitating cognitive impairment could be more likely to stop participating in the WLS than comparable non-football players whereas football players not suffering such early onset could be less likely to stop participating in the WLS than comparable non-football players.}

We posted our protocol that described our outcomes, matches and analysis plan on ArXiv on July 7, 2016 before estimating any treatment effects \cite{deshpande2016protocol}.

\subsubsection{Results}

Figure 1 Supplemental in the Supplementary Materials shows the distribution of the outcomes for the football players vs. the control groups after matching. For cognition, there was not a statistically significant effect of playing high school football vs. the combined control group with a 97.5\% confidence interval of reducing cognition by $0.18$ standard deviations to increasing cognition by $0.06$ standard deviations \footnote{This inference and the inferences that follow in this paragraph assume no unmeasured confounding.  See \cite{hasegawa2020causal} for sensitivity analyses.}  Judging by Cohen's criteria for effect sizes \cite{cohen1988statistical} (0.2 standard deviations for a small effect, 0.5 for a medium effect and 0.8 for a large effect), there is evidence that the effect of playing football on cognition was at most small.  For depression, playing football had a statistically significantly effect with a 97.5\% confidence interval of {\it{reducing}} depression by $0.25$ standard deviations to $0.02$ standard deviations.


\subsubsection{Value of Using the Protocol}

The results were not what I expected.  I thought we'd see a big negative effect of playing high school football and perhaps hoped (although it would have negative consequences for football players) to find such an effect, dreaming of getting mention in the {\it{New York Times}}.
However, we did not see any negative effect of playing high school football and if anything, we saw evidence of a beneficial effect of playing football on depression.  Our paper was rejected by the {\it{Journal of the American Medical Association (JAMA)}} (although published in its specialty journal {\it{JAMA Neurology}}).  Around the same time, {\it{JAMA}} published a paper ``Clinicopathological Evaluation of Chronic Traumatic Encephalopathy in Players of American Football'' which reported that brain damage was found in all but one of 111 brains donated by National Football League retirees and their concerned families.  The {\it{New York Times}} published an editorial ``The Scars from `Bell Ringing' Football Tackles'' based on the paper.  We wrote a letter to the {\it{New York Times}}:
\begin{quotation}
\noindent To the Editor:

``The Scars From ‘Bell-Ringing’ Football Tackles'' (editorial, 7/29) highlights a high rate of CTE [the type of brain damage football is hypothesized to cause] among former football players who donated their brains for study, and suggests safety concerns about not only pro but also high school football.  However, our research in {\it{JAMA Neurology}} finds among 1950s high school graduates, football players don’t have higher rates of later-life depression or impaired cognition than non-football players.   There are limitations to both our study (e.g., football has changed since the 1950s and youth sports is more professionalized) and the study cited by the editorial (e.g., self-selected and small sample of non-college/pro players).

Many people love playing football and playing promotes physical fitness, teamwork and recreation.  More research is needed into what amount of playing imposes substantial risks and what rule changes could reduce risks.

Former football players experiencing depression or other problems shouldn’t assume they have CTE and become fearful; they should seek help from mental health professionals and others as they very well may be just experiencing the usual challenges of life that most of us do.

\noindent Sameer Deshpande, Raiden Hasegawa, Dylan Small, Philadelphia, PA
\end{quotation}
The {\it{New York Times}} declined to publish our letter.

If we had not designed the study before looking at the outcome data (e.g., suppose we had just fit a regression of the outcome on playing football and the covariates), when we saw results that went against our hypotheses, I think we would have done additional analyses to see if the results might change.  And we might not have done some of these additional analyses had the initial results agreed with our hypotheses.  For a regression, investigation of diagnostics and the model fit is part of good practice.  But a bias can arise if when the results go against what we expected, we investigate more than we do when the results are in line with what we expected \cite{buja2014discussion}.  Such selective investigation is hard not to do.  Dante wrote in his {\it{Divine Comedy}} \cite{alighieri1320} of St. Thomas Aquinas cautioning him on entering paradise:
\begin{quote}
Opinion {\textendash} hasty {\textendash} often can incline to the wrong side, and then affection for one's own opinion binds, confines the mind.
\end{quote}
Designing the study and writing the protocol before fitting a model to estimate a treatment effect protected us against this confirmation bias.  It made our study more objective, more reliable and more transparent which promotes scientific progress.  When I've made this argument about the value of protocols for observational studies, people have given me the criticism that unlike in randomized trials, in a retrospective observational study in which the outcomes are already available when the protocol is posted, an investigator could have cheated and analyzed the outcomes before posting the protocol.  This is true {\textendash} however, the goal of an observational study protocol is not to protect against dishonest investigators but to aid honest investigators to do good science.

\subsection{Confiscating Guns Used in Domestic Violence}

A variety of laws in the U.S. restrict the possession of guns by individuals convicted of domestic violence or under a restraining order.
Pennsylvania has a law going further that restricts possession of a gun before conviction.  The law requires a police officer called to a domestic violence scene to remove a gun when making an arrest if the gun was used in the commission of the alleged domestic violence; furthermore,
the officer may exercise their judgment and remove the gun even if not making an arrest.  In \cite{small2019after}, we studied whether this law works to prevent the escalation of domestic violence using data from 2013 provided by the Philadelphia Police Department.  In practice, a gun was not always {\textendash} among 220 such domestic violence incidents involving a gun in 2013, the gun was removed in only 52 (24\% ).  We asked, does removing the gun reduce the incidence of future calls about domestic violence to the police?

\subsubsection{Protocol}

There were 68 potential variables available for matching that came from a form a police officer must complete when responding to a domestic violence incident.  After discussing with my collaborators, we decided to match on the variables in Table \ref{balance.table.gun}.  Before matching, there were many substantial differences between the gun removed (treated) and gun not removed (control) groups such as the gun was less likely to be removed if the suspect fled the scene and more likely to be removed if the suspect was arrested at the scene.  After matching,
my collaborators felt the balance was acceptable with all absolute standardized differences less than $0.2$, a criteria for acceptable balance \cite{silber2001multivariate}.  In our protocol, we provided the balance table for the match, specified our primary outcome as future domestic violence calls (which might or might not involve a gun) following a domestic violence call that involved a gun, and specified our planned analysis as robust Poisson regression with inverse probability weighting to estimate the treatment on treated effect.

\begin{table}[ht]
\centering
\caption{Covariate balance, before and after matching, in the study on the effect of confiscating guns used in domestic violence.  Because matched sets vary in size,
means are weighted to reflect frequencies in the treated group.  The standardized difference is the difference
in means divided by the standard deviation prior to matching.  Standardized differences above
0.2 are in \textbf{bold}.}
\begin{tabular}{l|rrr|rr}
 \hline
 & \multicolumn{3}{|c}{Weighted Mean} & \multicolumn{2}{|c}{Standardized Difference} \\ \hline
& Treated & Matched  & All  & After & Before  \\
& & controls & controls & matching & matching \\
 \hline
Sample size & 52 & 148 & 168 & 52 vs. 148 & 52 vs. 168 \\ \hline
Offender Covariate & & & &  \\
\ Male & 0.87 & 0.89 & 0.88 & -0.08 & -0.05 \\
\ Black & 0.69 & 0.77 & 0.80 & -0.19 & -0.26 \\
\ Age (in years) & 36.83 & 35.17 & 32.06 & 0.14 & \textbf{0.41} \\
\ History of substance abuse & 0.13 & 0.18 & 0.18 & -0.13 & -0.12 \\
\ Under court supervision & 0.02 & 0.04 & 0.10 & -0.07 & \textbf{--0.35} \\
\ History of domestic violence & 0.27 & 0.27 & 0.42 & 0.00 & \textbf{--0.32} \\
\ History of domestic violence & 0.15 & 0.20 & 0.26 & -0.11 & \textbf{--0.27} \\
\ \ reported to police & & & &  \\
\ Fled scene & 0.37 & 0.36 & 0.80 & 0.00 & \textbf{--0.99} \\
\ Arrested at scene & 0.75 & 0.74 & 0.22 & 0.03 & \textbf{1.24} \\
Victim Covariate & & & & \\
\ Female & 0.83 & 0.86 & 0.89 & -0.10 & -0.19 \\
\ Age (in years) & 33.52 & 32.22 & 30.19 & 0.12 & \textbf{0.32} \\
\ Emotional reaction (range: 0-4) & 2.02 & 1.98 & 1.49 & 0.03 & \textbf{0.63} \\
  \hline
\end{tabular}
\label{balance.table.gun}
\end{table}%

\subsubsection{Results}

Using our prespecified analysis plan, we estimated that removing a gun {\it{increased}} the rate of future domestic violence incident calls 5.58  times (95\% confidence interval (1.69, 18.39); $p$-value $<0.01$).\footnote{These inferences are based on the match presented in \cite{yu2021information}}

\subsubsection{Value of Using the Protocol}

My collaborator Susan Sorenson is a violence prevention researcher and an activist on domestic violence prevention.  I e-mailed Susan to ask if she had any gun violence prevention projects I might be helpful for on November 6, 2017 after the Sutherland Spring church mass shooting occurred the previous day in which a shooter killed 26 people.  Susan suggested the project we worked on with the hope that it would demonstrate that Pennsylvania's law was effective in preventing domestic violence.  Both of us were sorely disappointed in the results which suggested the law increased future domestic violence calls.  Post hoc, we thought of certain explanations for the results which were consistent with the law being effective such as that domestic violence victims might feel safer and less fearful of retribution without the gun in the home and more comfortable to call for assistance.  However, a less positive explanation is that if an abuser no longer has access to a gun, the abuser may resort to more visible domination strategies {\textendash} threatening more often and more aggressively {\textendash} leading to more domestic violence calls.  Overall, the results can hardly be used to advocate for more widespread use of gun removal laws to prevent domestic violence.

Suppose that instead of writing a protocol and thinking hard about which of the 68 variables on the police form to match on, we had chosen some variables initially and fit a preliminary regression model with the intention to check and revise the model after seeing the initial model fit.  This is a suggested and sensible strategy for using regression models for many data analysis tasks (e.g., see Display 9.9 in \cite{ramsey2002statistical}), but it is problematic for making causal inferences.  After we saw the result that the Pennsylvania law increased domestic violence calls instead of our hoped for reduction, we might have checked the model more intensely and made more revisions to it than if the results had been in our hoped for direction.  For example, we might have revised the model by including different variables or changing the coding of variables.  Examples of issues with coding the variables that we encountered were the following: for race/ethnicity, whether to have a separate categories for Asian, white, black and Hispanic (we decided to just have black/not black because of low counts if we used all categories) and how to handle a small number of missing values for race or sex (we decided to impute them to the most common category).  Our choices could be argued with, but having made them blinded before seeing the treatment effect estimates increases our confidence that the results are reliable.  

For some of the implementation or analytical choices that must be made in a study, it is reasonable to conduct a stability analysis to examine the effect of the choices.  But it is desirable to minimize the need for such stability analyses to aid critical evaluation of the study \cite{rosenbaum1999choice,rosenbaum2010design}.  Suppose 250 stability analyses are done in an analysis relying heavily on complex analytic models and 240 of them support the original findings.  It would be time consuming for subject matter experts to critically evaluate whether the choices made in the 10 stability analyses that do not support the original findings are reasonable.  But if say two stability analyses are done and one supports the original findings, it is more manageable for a subject matter expert to evaluate the choice made in the one stability analysis that does not support the original findings.  By thinking through the analytic choices before seeing the treatment effect estimates and writing them in a protocol, we can restrict stability analyses to only the choices that seem more important.

\subsection{Mountaintop Mining and Low Birthweight}

Surface mining has become a significant method of coal mining in the Central Appalachian region of  the  eastern  United  States  alongside the  traditional  underground  mining.  A substantial proportion of the surface mining is mountaintop removal mining (MRM), where the tops of mountains are blasted off to reach the coal seams that lie underneath. Much of the debris is then dumped into surrounding valleys. MRM is controversial {\textendash} coal companies claim economic benefits and improved safety for workers compared to underground mining, while Central Appalachia residents have raised concerns about its environmental and health effects \cite{house2009something}.
In 2016, under the Obama administration, the U.S. Department of the Interior's Office of Surface Mining Reclamation and Enforcement (OSMRE) contracted with the National Academies of Science, Engineering and Medicine (the National Academies) to review the public health risks of MRM.   The National Academies assembled a panel of 11 scientists and experts.  The panel was about five months into its work and was hours away from holding a public meeting to hear from coalfield residents in Hazard, Kentucky when word arrived from the OSMRE, now under the control of the Trump administration, that the study must be stopped immediately \cite{ward2017}. I heard about this when serving on a different National Academies panel (on the generational health effects of serving in the Gulf War) from a frustrated National Academies' staff member who had worked with the MRM panel until it was disbanded.
I realized that birth certificate data might be useful for looking at the effect of surface mining on low birthweight and set out with collaborators to design an observational study \cite{small2021surface}. We chose our exposure to be surface mining, rather than MRM specifically, because an estimate of surface mining activity was was available using Google Earth Engine and a method developed by \cite{pericak2018mapping}.  We used a control group design with a pretest and a posttest in which the pretest period was 1977–1989, a period of low surface mining activity and three posttest periods were considered, 1990–1998, 1999–2011 and 2012–2017,  Surface mining in Central Appalachia increased after 1989, partly because of the Clean Air Act Amendments of 1990 which made surface mining in Appalachia more financially attractive \cite{hendryx2016unintended}.  In the late 1990s, surface mining increased further because of the development of bigger equipment that enabled mining at scale \cite{ramani2012surface}.  In the period 2012-2017, surface mining of coal declined as energy demand shifted away from coal toward natural gas. We chose for our primary analysis the posttest period of 1999-2011, a period of high and sustained surface mining, and considered the other two posttest periods in secondary analyses.  Our treatment group was Central Appalachian counties with a high amount of surface mining between 1999-2011 and our control group was other counties in the four states in which MRM takes place {\textendash} Kentucky, Tennessee, Virginia and West Virginia. We constructed a matched control group by matching treated counties to control counties on socioeconomic characteristics and maternal smoking rates.  Our planned analysis was a difference-in-differences type analysis in which we fit a logistic regression model of the outcome on dummy variables for each county, year dummy variables, the individual covariates and a variable for the amount of surface mining in a treated county in the post period.  We also planned a sensitivity analysis for unmeasured confounding if the effect of high surface mining was significant in our primary analysis.  One of my collaborators raised the issue that counties that border surface mining counties might have spillover effects from the mining via air or water pollution.  To assess whether including border counties would affect our results, we specified a stability analysis of using a different set of matched control counties that excluded border counties.  After matching and planning the analysis, but before conducting the analysis, we posted our protocol on ArXiv \cite{small2020protocol}.

\subsubsection{Results and Interpretation}

We found not significant results for our primary analysis posttest period of 1999-2011 as well as our secondary analysis posttest period of 1990-1998 but a significant increase in low birthweight for our secondary analysis posttest period of 2012-2017.  Specifically, compared to the pre-period of low surface mining from 1977 to 1989, for the primary analysis posttest period of 1999–2011, we estimated a relative  increase in low birth  weight in  surface  mining  counties  compared  to  matched  control  counties  that  was  not  statistically  significant  (odds  ratio for a 5 percentage point increase in area disturbed by surface mining: 1.07,  95\%  confidence  interval: 0.96, 1.20);   for  the  secondary  analysis  posttest  period  of 1990–1998,  there was  not a relative increase (odds  ratio:  0.98, 95\%  confidence  interval: 0.74, 1.13) and for the secondary analysis posttest period of 2012–2017, there was a statistically significant relative  increase (odds  ratio: 1.28, 95\%  confidence  interval: 1.08, 1.50).  The results were similar for the stability analysis that only considered control counties that do not border surface mining counties.

In planning our analysis, we expected that if there was an effect of surface mining on low birthweight, it would be strongest when surface mining activity was highest.  This is why we chose 1999-2011, a period of high and sustained surface mining activity, as our primary analysis posttest period.  After finding stronger evidence for an effect in the secondary analysis posttest period of 2012-2017 when surface mining had declined, we developed the following post hoc explanation: it takes time for the impact of an area's hydrology (e.g., valley fills and changes to hydrological structures from blasting) to show up in the water supplies that affect human health.

To investigate the possibility of a delayed effect of surface mining, we did a post hoc analysis. We fit the following change point model :  a logistic regression model of low birthweight on county and year dummy  variables, the individual covariates  and  a  dummy variable for whether the observation is in a surface mining county (as compared to a matched control county) and the year is $k$  or  later,  where $k$ could range from 1990 to 2017 or be  $\infty$ (which would mean surface mining had no effect). We tested the null hypothesis that $k=\infty$, i.e., that surface mining  had no effect, using the parametric bootstrap.  The $p$-value was $<0.01$, evidence that surface mining had an effect and the estimated $k$ (time of onset of an effect) was 2005 with a 95\% confidence interval of 2000 to 2013.

\subsubsection{Value of Protocol}

I am a registered Democrat and I was perhaps partly motivated to do this study to show that by halting the National Academies' study, the Trump Administration was harming health to serve its political interests.  From this perspective, the primary analysis finding of no significant effect of surface mining was disappointing to me.  Had we not clearly demarcated the primary, secondary and post hoc analyses, we might have presented the results in a way, even if unintentionally, that more strongly confirmed our prior hypothesis of surface mining having a harmful effect than is appropriate based on the data. For example, we might have presented as our main analysis the post hoc change point model that showed that there was evidence that surface mining had an effect ($p<0.01$).\footnote{The German philosopher Arthur Schopenhauer observed \cite{schopenhauer2011world}
\begin{quote}
An adopted hypothesis gives us lynx-eyes for everything that confirms it and makes us blind to everything that contradicts it.
\end{quote}}
Instead, because our protocol clearly demarcated the time period 1999-2011 as the primary analysis and the time periods 1990-1998 and 2012-2017 as secondary analyses, in the discussion of our paper, we made clear that our primary analysis was not significant and only a secondary analysis was significant \cite{small2021surface}:
\begin{quote}
In our primary analysis of the period 1999–2011 and our secondary analysis of the period 1990–1998, we did not find evidence of an effect of surface mining on low birthweight; however, in our secondary analysis  of the years 2012–2017, we did find evidence that surface mining was associated with low birthweight.
\end{quote}
This appropriately tempers how much the data supports our post hoc theory that there was a delayed effect of the surface mining that started ramping up in the 1990s and became heaviest in 1999-2011.  See Section \ref{open.problems} for discussion of open problems related to how to present secondary analysis findings when the primary analysis is not significant but a secondary analysis is.

\section{Common Features of Good Observational Study Protocols and Good Randomized Trial Protocols}
\label{sec: common features}

The goals of an observational study protocol and randomized trial protocol are similar, including to  \cite{chan2013spirit,wang2022reproducibility,wang2022harmonized}
\begin{enumerate}
\item Help the investigators think through design and analysis choices before seeing the results in order to protect themselves from confirmation bias.
\item Facilitate review of the study by enabling reviewers to assess potential biases of these choices and to track any changes in the actual analysis from the pre-specified methods;
\item Facilitate reproducibility of the study.
\end{enumerate}

Much of what makes a good observational study protocol (e.g., \cite{wang2022harmonized}) is what makes a good randomized trial protocol (e.g.,  \cite{chan2013spirit}).  Common elements include an unambiguous, complete and transparent description of the following: inclusion and exclusion criteria for subjects; the exposure; the primary and secondary outcome(s); the covariates that will be adjusted for; the statistical methods that will be used to compare the treatment and control group, including which analysis will be considered the primary analysis, what stability analyses will be conducted, how missing data will be handled, subgroups that will be considered and how multiplicity will be handled; human subject protections; how the data will be managed to protect subject confidentiality; rationale and background for the study; funding sources for the study and potential conflicts of interest.  There are some differences in what should be described in observational study protocols compared to randomized trial protocols due to the nature of the studies.  For example, in observational studies, the duration of the treatment is not controlled by the investigator and so decisions must sometimes be made about how to define the duration from the available data, e.g., in a pharmacoepidemiologic study, decisions must be made regarding how to handle early refills of a drug or conversely short gaps in between prescriptions \cite{wang2022harmonized}.  In observational studies, the investigator cannot choose to blind subjects or evaluators to the treatment (although sometimes this happens naturally) but in randomized trials, blinding can be done and the protocol should describe if and how blinding will be done \cite{chan2013spirit}.

\subsection{Making a Protocol Adaptive}

Investigators may be concerned that writing down too strict a protocol will limit their ability to learn from the data through exploratory data analysis and to make analytic choices that maximize power.  The power of a study can depend on features of the analytic plan such as how much a test statistic emphasizes tail observations, how outliers are handled, and how multiple outcomes and subgroups are incorporated into the analysis.  The best choices for these features often depend on aspects of the population about which one is uncertain before obtaining data.   Ideally one would like to adapt the analytic plan to the data but if there is no protocol which prespecifies the analysis, the researcher’s biases (conscious or subconscious) can influence the results \cite{chan2004empirical}.  An adaptive protocol allows for adapting to the data while protecting against researcher bias.

One approach to making a protocol adaptive is to specify several (or an infinite family of) analyses that will be considered and account for considering multiple analyses in making inferences \cite{rosenbaum2012exact,balzer2016adaptive,rosenbaum2017adaptive}.  For example, consider testing the null hypothesis of no treatment effect in a matched pair study where we would like the Type I error rate to be 0.05.  Different tests could be used, e.g., Wilcoxon's signed rank test, Brown's test or Noether's test.  We could use two or more of these tests simultaneously but would have to account for the multiplicity of hypothesis testing.  Because these tests are highly correlated, the Bonferroni correction is conservative and more specialized methods of correcting for multiplicity are desirable.  For example, \cite{rosenbaum2012testing,heng2021increasing} develop approaches for using two tests together that asymptotically control the Type I error rate by considering the asymptotic distribution of the maximum of the test statistics, and \cite{rosenbaum2012exact} develops an approach for using specifically Brown's test and Noether's test together that exactly controls the Type I error rate by making use of the two tests' exact joint distribution.\footnote{This adaptive test is implemented in the \texttt{adaptive.noether.brown} function in the \texttt{SensitivityCaseControl} R package.}  Another example of an adaptive protocol based on accounting for the multiplicity in considering multiple analyses when considering subgroups \ref{lee2018powerful} will be discussed in Section \ref{sec: subgroups}.  

Another approach to making a proposal adaptive is to randomly split the sample and choose the analysis based on the first part of the sample and make inferences using the chosen analysis on the second part of the sample \cite{cox1975note,heller2009split}.  This approach will be discussed more for selecting outcomes in Section \ref{sec: multiple outcomes} and selecting subgroups in Section \ref{sec: subgroups}.

A third approach to making a protocol adaptive is to use an aspect of the data to adapt the protocol which is not affected by treatment assignment under the null hypothesis of no treatment effect, so that the randomization inference after adapting the data analysis plan is the same as if no adapting had been done.\footnote{See \cite{rosenbaum2002obsstudies,rosenbaum2002covariance} for discussion of randomization inference for observational studies.  Fisher \cite{fisher1935statistical} described randomization as the ``reasoned basis for inference'' in randomized experiments.  Randomization inference is valid for observational studies under the assumption of no unmeasured confounding.} For example, consider outliers.  If an outlier data point is deemed not plausible (e.g., because it contains data entry or processing or measurement errors, or because it comes from a different population than the one the investigator wants to study), then it may be sensible to remove the point, but if the point just represents natural variation, removing it may create bias.  Deciding what are the outliers requires looking at the outcome data, which is usually not done until after writing the protocol.  Furthermore, making decisions about whether to remove an outlier requires judgment that could be biased if one uses the outcome data to look at how the outlier affects the treatment effect estimate.  A way around these problems is to say in the protocol that when deciding whether to remove an outlier, we will only look at the outcome data alone, and not look at the joint outcome and treatment data.  Then under the sharp null hypothesis of no treatment effect on an individual level, the decision about whether to remove an outlier would be the same regardless of the treatment assignment so that usual randomization inference can still be used \cite{zhang2021method}.  For example, in studying the effects of social distancing on COVID-19 cases using county-level data, \cite{zhang2021method} used this approach to discover that several counties with outlying high COVID-19 rates had large prison outbreaks of COVID-19.  These counties were removed from the analysis because individuals in prisons had no choice as to whether to social distance, so cases due to prison outbreaks were outside the scope of the intended analysis.  A limitation to this approach to removing outliers is that it can only strictly be used to test the null hypothesis of no treatment effect.\footnote{It can be used to test a pre-specified additive treatment effect if the investigator looks at the potential outcome under control assuming  that additive treatment effect, i.e., the outcome for control subject and the outcome minus the treatment for treated subjects, {\it{without}} looking at the treatment status \cite{zhang2021method}}.  We will consider another instance of the approach of looking at an aspect of the data which is not affected by treatment assignment under the null hypothesis of no treatment effect for choosing subgroups \cite{hsu2013effect,hsu2015strong,lee2018discovering}
in Section \ref{effect.modification.secondary}.

\subsection{Post Hoc Analyses with a Protocol}

In carrying out a study, it is common that considerations arise that we did not expect when writing the protocol.  For example, in the mountaintop mining study, our finding that the mining only had a significant effect for the last of the three periods considered in the protocol (2012-2017) motivated us to do a change-point analysis of what exact year the effect started.  In the football study, we planned a sensitivity analysis on the condition that we found a significant effect.  But post hoc, we realized that a finding of equivalence (football players being no different than non-football players within a specified equivalence margin) was also of interest and we post hoc specified an equivalence margin and did a post hoc sensitivity analysis \cite{hasegawa2020causal}.\footnote{One approach to simultaneously testing for a difference and equivalence is the three-sided test of \cite{goeman2010three} for which sensitivity analysis can also be conducted 
\cite{pimentel2015large}.}

Some post hoc analyses can be avoided by designing better protocols, but even the most experienced and thoughtful investigators sometimes think of additional analyses they wish they had put in the protocol \cite{cox2016design}.  These additional analyses can be presented and labeled as post hoc analyses, which are exploratory, whereas the prespecified analyses in the protocol are confirmatory.\footnote{Sometimes it is necessary to make changes to the protocol rather than just add post hoc analyses.  For example, suppose it was discovered after starting the analysis that an outcome was not measured in the way the researcher thought it was {\textendash} then it might be necessary to alter how the outcome is defined.  It is important to have a way of keeping track of such changes to the protocol.  This can be done in a preprint server like ArXiv or Zenodo by posting updated versions of the protocol that contain a table detailing what has changed from the initial version \cite{wang2022harmonized}; the earlier versions of the protocol remain accessible.}  Tukey talked about the importance of both confirmatory and exploratory analyses \cite{tukey1980we}:
\begin{quote}
We do not dare either give up exploratory data analysis or make it our sole interest...We often forget how science and engineering function.  Ideas come from previous exploration more often than from lightning strokes.  Important questions can demand the most careful planning for confirmatory analysis.
\end{quote}


While we should always carry out the analyses specified in the protocol, we should also always just look at the data through exploratory data analysis techniques such as plots \cite{tukey1977eda}.  Janet Wittes \cite{wittes2018} quoted William Cochran's advice, ``Don't let arcane theory prevent you from looking at actual data.''  Colin Mallows remarked \cite{rosenbaum2020modern}, ``The most robust method I know is to look at the data.'' \cite{rosenbaum2022new} discusses plots that are useful for looking at treatment-minus-control matched pair differences in observational studies and randomized trials. \cite{buja2009statistical} discusses statistical inference techniques for exploratory data analysis that assign $p$-values to visual discoveries.\footnote{One method, the line-up protocol, consists of generating, say, 19 null plots, inserting the plot of the real data in a random location among the null plots and asking the human viewer to single out one of the 20 plots as most different from the others. If the viewer chooses the plot of the real data, then the discovery can be assigned a $p$-value of 0.05 $(=1/20)$  \cite{buja2009statistical}.}  The protocol could specify planned exploratory data analyses.  Such prespecification helps to reduce bias from multiplicity if it is uncontrolled for and provides a basis for controlling for the multiplicity by using multiple testing techniques.  However, if the exploration suggests additional exploratory analyses, e.g., additional plots, these should be pursued \cite{diaconis1981magical}.


\section{Different Considerations in Observational Study Protocols vs. Randomized Trial Protocols}
\label{sec: different.considerations}

Although much of what makes a good protocol for an observational study is what makes a good protocol for a randomized trial, there are some differences. We discuss three in this section.

\subsection{Protocol Before Treatment Assignment or After Treatment Assignment}
\label{protocol.before.vs.after}

In an RCT, the protocol is typically written before subjects are randomly assigned to treatment \cite{chan2013spirit}.  In a matched observational study, matching treated and control subjects (imperfectly) takes the place of random assignment, but protocols are typically written after matching \cite{rubin2007design,rosenbaum2020modern}.  A potential disadvantage of writing a protocol after matching is researcher hope bias from judgment about which covariates to match on.  We describe this potential bias below.

In an observational study, there is frequently tension between designing a study that attempts to rigorously control for all possible confounding variables to avoid bias and designing a study that has a chance of providing informative results.  In full matching, matching on a variable that is highly out of balance between the treatment and control group drives up the variance of the estimated treatment effect.  This is analogous to how in regression, controlling for a variable that is highly correlated with the treatment drives up the variance of the estimated treatment effect due to multicollinearity.  There are several types of variables for which it can be better not to match under certain circumstances such as instrumental or near instrumental variables \cite{myers2011effects,wooldridge2016should,pimentel2016constructed}, post-treatment variables \cite{rosenbaum1984consequences} or proxies for the treatment \cite{breslow1980statistical,marsh2002removal}.  Deciding whether it is better or not to match on such a variable requires judgment about the potential bias from not matching which involves subject matter considerations and cannot be automated.  We illustrate this in the context of potential instrumental variables (IVs).

An IV is a variable that is associated with the exposure but not with unmeasured confounders and that does not directly affect the outcome (i.e., affects the outcome only through its association with the exposure) \cite{baiocchi2014instrumental,imbens2014instrumental}.  An IV should typically not be matched on.  An IV provides useful variation in the treatment that is free of unmeasured confounders, but matching on an IV effectively conditions on it, leaving more of the variation in the treatment that is due to unmeasured confounders and consequently amplifying bias \cite{rubin1997estimating,brookhart2006variable,austin2007comparison,myers2011effects,wooldridge2016should,pimentel2016constructed}.  In simulations, \cite{myers2011effects} found that not matching continued to be better for near-IVs that are slightly associated with unmeasured confounders (so are confounders themselves) but matching was better for not-so-near IVs that had a moderate association with unmeasured confounders.  They concluded,
\begin{quote}
The results indicate that effect estimates which are conditional on a perfect IV or near-IV may have
larger bias and variance than the unconditional estimate. However, in most scenarios considered, the increases in
error due to conditioning were small compared with the total estimation error. In these cases, minimizing unmeasured
confounding should be the priority when selecting variables for adjustment, even at the risk of conditioning on IVs.
\end{quote}
In the Supplementary Materials, we expanded \cite{myers2011effects}'s simulations to include variables that are proxies for the treatment (highly associated with the treatment). \cite{myers2011effects}'s original simulations suggested that for variables that are not close proxies for the treatment, it is sensible to err on the side of matching on a variable if it might be associated with unmeasured confounders.  However, the additional simulations we did suggest that when a variable is a close proxy for the treatment, it is often better to not match on it.

We considered two possible IVs for the gun study {\textendash} whether the suspect fled the scene ({\it{fled}}) and whether the suspect was arrested at the scene ({\it{arrested}}).  Whether the suspect fled the scene affects whether the gun was removed because typically if the suspect fled the scene before the police arrived, the suspect  would have taken their gun with them so that the gun cannot be confiscated (when the suspect fled the scene, the gun was confiscated only 12\% of the time but it was confiscated 50\% of the time when the suspect did not flee).  Also, whether the police officer arrested the suspect at the scene has a big impact on whether the gun was confiscated (the gun was confiscated 51\% of the time when the suspect was arrested but only 9\% of the time when the suspected was not arrested).  Not matching on {\it{fled}} and {\it{arrested}} would decrease the variance of the treatment effect estimate {\textendash} the effective sample size \cite{hansen2004full} from not matching vs. matching increases to 60 pairs from 50 pairs
However, it is questionable whether a suspect fleeing the scene or whether an arrest was made are independent of future violence risk (an unmeasured confounder), e.g., a suspect fleeing the scene might suggest a lack of remorse and increase the risk of future violence.  Thus, {\it{fled}} and {\it{arrested}} may not be valid or even near-valid IVs.  Figure 3 in the supplementary
materials shows whether it is better to match or not match on these variables in terms of root mean squared error for various values of parameters describing how invalid the variables are as IVs.  Considering the moderate gain from not matching even when they are perfectly valid IVs and the possibility that they are considerably invalid, we chose to match on them.  

The effective sample size from matching on all variables was 50 and from not matching on the two proposed IVs {\it{fled}} and {\it{arrested}} was 60 so that there was only a moderate gain from not matching.  But suppose that the effective sample size from matching on all variables was 20.  Then there would be a little power from matching on all variables without an enormous effect size and the only hope for much power would be not to match on the proposed IVs.  One approach would be to say that since we are concerned that these proposed IVs are actually confounders and so invalid IVs, we think it is still best to match and either carry out the study with the admitted limitation that it may be underpowered, or not carry out the study further because of the lack of power.  But some researchers might try to be optimistic and not match, hoping for the best.  This can cause {\it{researcher hope bias}} {\textendash} bias from not matching on potential confounders because the researcher hopes they are not true confounders to give the study power rather than realistically thinks they are not true confounders.

One way to reduce researcher hope bias is to write the protocol before matching.  This will prevent a researcher's judgment about whether it is better or not to match for a variable from being clouded by seeing how matching vs. not matching on the variable affects the effective sample size.  For a protocol written before matching, the variables to match on could either be chosen by (i) the traditional approach of having a subject matter expert(s) think about what variables are confounders without examining the data or (ii) automated, data driven methods for choosing variables \cite{schneeweiss2009high,patrick2011implications}.\footnote{\cite{patrick2011implications} suggest the following as a middle ground between having subject matter experts choose which covariates to match on without looking at the data and a purely data-driven approach: use a data-driven approach to identify a pool of potential variables to match on and then the researcher could identify and exclude potentially problematic variables in that pool (e.g., variables that are IVs or not pre-treatment).  However, if the researcher examines the joint covariate-treatment data in making decisions about which variables to exclude, e.g., by looking at an initial match using all variables in the pool of variables identified by the data driven approach, then researcher hope bias could creep in.}  After specifying what variables will be matched on or a method for how those variables will be chosen, a protocol written before matching should explain how the match will be conducted (e.g., what distance metric will be used such as propensity scores or Mahalanobis, or whether pair matching or mathcing full will be used).  The protocol could specify that different matching methods will be tried and the one that maximizes balance in some sense (e.g., \cite{harder2010propensity}) will be used \cite{cafri2018mitigating}.  

Writing an observational study protocol before matching makes it analogous to a randomized trial protocol.  However, there is a central difference between a randomized trial and an observational study.  In a randomized trial, while adjusting for covariates can improve the efficiency, it is not necessary to produce unbiased results.  But in an observational study, figuring out which covariates to adjust for is critical and the process of looking at the data through matching can provide insights.  \cite{yu2021information} suggest constructing an initial match with the variables that a subject matter expert(s) thinks are important, then screening a potentially large list of other variables to see which ones are substantially imbalanced between the treatment and (initial) matched control groups and then thinking about which of these imbalanced variables we should also match on.  In an example looking at the effect of hormone replacement therapy, this approach identified a potentially important variable, history of breast cancer, that had not been adjusted for in previous studies \cite{yu2021information}.  History of breast cancer could also have been identified as an important variable to match on in a protocol before matching that specified the matching variables would be chosen in an automated, data driven way. But a concern with a purely automated approach is that thinking is required in making decisions about whether to adjust for a variable, such as whether to adjust for a variable that might or might not be an IV (as discussed above, this depends on judgment about how associated the potential IV is with unmeasured confounders) or a variable that might look like a pretreatment variable but not be.\footnote{An example of a variable that looks like a pretreatment variable but is not is given by \cite{niknam2018adjustment}.  Various studies that rank US hospital performance at treating acute myocardial infarction
(AMI) adjust for a diagnostic code of atherosclerosis. Some degree of atherosclerosis is common in older Americans. The actual disease of atherosclerosis is certainly a covariate when treating AMI, so one is inclined to include it in adjustments based on its name alone.  But \cite{niknam2018adjustment} show that the notation of atherosclerosis in a Medicare claim for AMI commonly indicates that a percutaneous coronary intervention has been performed, so the notation of atherosclerosis in a Medicare claim is an outcome of a particular treatment.  \cite{niknam2018adjustment} go on to show that adjustments should not be made for atherosclerosis when evaluating hospital performance at treating AMI.}  Thinking about the status of many variables is difficult but thinking about a few variables that have been identified after an initial match is more manageable.  

Another advantage of exploring an initial match before choosing the variables for a final match is that the initial match can generate ideas for additional variables to construct that might reduce unmeasured confounding.  In the gun study, in addition to the police officer filling out a structured form for each incident, the officer provided a text note containing further information.  The text notes could be used to define additional potential confounding variables beyond those in the structured form, but reading through all the text notes would be time consuming.  To make looking at text or other qualitative data feasible in a reasonable amount of time, \cite{rosenbaum2001matching} proposes looking at the qualitative data in a few matched sets in detail. \cite{yu2021optimal} develop an optimal matching approach (implemented in the \texttt{R} package \texttt{thickmatch}) for choosing which matched sets to look at {\textendash} the approach balances the quantitative matching variables on the structured form (i.e., the variables in Table \ref{balance.table.gun}) and finds a subset of exceptionally close pairs on all of the variables for which the text notes can be compared between the treated and control units.  This comparison might suggest potential confounding variables that could be extracted from the remaining text notes.  For the gun study, \cite{yu2021optimal} considered the text notes in five exceptionally close pairs and noted differences in whether the suspected offender threatened to kill or shoot someone (e.g., in one incident, the suspected offender shot the complainant in the chest and in another incident, the suspected offender shot the front door) and in whether children were involved (e.g., in one incident, the text note said ``Comp [Complainant] states that the suspected offender was intoxicated and came to her house (apt) and pointed a black hand gun at her and took their 3 month baby out of her apt and drove away.'')  Whether the suspected offender threatened to shoot or kill someone and whether there were children involved are plausible considerations when deciding whether to confiscate a gun, but these are not reflected in the covariates on the police form.  We created two new binary covariates based on these considerations using the \texttt{grep} function in \texttt{R} to search the text for keywords.  The keywords were permitted to differ by one letter from the intended keyword. The first covariate was 1 if either the word ``kill'' appeared or the phrase ``shoot you'' appeared.   The second covariate was 1 if any of the following words appeared: child, baby, daughter, son, dad, mom, father, mother, daddy, mommy, pregnant; otherwise, the second covariate was zero.  We matched on these two new text covariates along with the original 12 matching variables.  Matching for the two new text covariates slightly increased the point estimate but the confidence intervals and $P$-values were similar to those of the original match, and we concluded \cite{yu2021optimal}:
\begin{quote}
In summary, close reading of the narratives for a few closely matched pairs raised a concern that we have might missed important sources of confounding.  We generated new covariates to capture the potential sources of confounding identified from reading the narratives and we matched on these covariates in addition to the covariates included in our initial match.  With this new match, we obtained similar inferences about the effect of gun removal as with our original match, enhancing our confidence in the inferences from our original match.
\end{quote}

Another advantage of exploring an initial match before finalizing the protocol is the exploration can suggest quasi-experimental devices to add that strengthen the design of the observational study.  In the football study, after looking at the match and being asked if there were any biases to worry about, one of our collaborators commented,  ``depressed poets don't play football.'' This inspired us (whether the comment is correct or not) to design multiple control groups\footnote{(i) men who played sports but only non-collision sports and (ii) men who did not play a sport at all} to check for potential unmeasured confounding from a person's interest in playing sports.  

In summary, there are some considerations that argue for specifying a protocol before matching and some for after matching.  In some sense, the issue of whether to specify a protocol before matching to avoid understating the uncertainty in a study that can come from specifying it after matching is analogous to the issue of whether to specify a protocol before looking at the outcome data to avoid the bias that can come from specifying it after looking at the outcome data.  In both cases specifying ``before'' minimizes certain types of misleading inferences but may sacrifice information.
For the latter issue of specifying the protocol before vs. after looking at the outcome data, we have argued for specifying before to avoid bias, but for the former issue of specifying before vs. after the match, we have taken a less strong position.  For specifying before vs. after the match, the potential of researcher hope bias (understatement of uncertainty from specifying a protocol after matching because the researcher hopes it will be feasible to learn something useful from the data) needs to be traded off against the information one can learn about unmeasured confounding from specifying a protocol after matching.  One potential way to reduce researcher hope bias is to before matching, specify the list of covariates that are thought to be important to match on and provide reasons in the final protocol for any changes from this initial list \cite{cafri2018mitigating}.  It would be useful if future work could provide other ways of reducing researcher hope bias.


\subsection{Multiple Outcomes}
\label{sec: multiple outcomes}

In the football study, we considered as primary outcomes one measure of later life mental health (the CES-D depression score at age 65) and one measure of later life cognition (a combination of delayed word recall and letter fluency at age 65).  There were many possible outcomes we could have considered, e.g., for mental health, there were measures of depression, hostility, anxiety, anger and heavy drinking and for cognition, there were six tests {\textendash} immediate word recall (after 10 words are read, recall as many as can), delayed word recall (12 minutes after the immediate word recall, recall as many words as can), digit ordering (participants started with a set of three digits read to them at one-second intervals and had to order them from smallest to largest; if correctly ordered, four digits are read...continue until participant cannot correctly order the digits or the upper limit of 8 is reached), similarities (subjects were asked how two things are alike, e.g. an orange and a banana,
and scores of 0, 1, or 2 were assigned based on the correctness and categorical abstractness of the response, e.g.,
``They are both fruits'' gets a score of 2, while ``They both have skins'' gets a score of 1); letter fluency (person is asked to name as many words as start with a randomly chosen letter among the letters ``F'' and ``L''); and category fluency (participants were asked to name either as many foods or as many animals as possible in one minute).  We could also have considered each of these tests at ages 53, 65 or 72.  Additionally, we could have considered sub-aspects of each test, e.g., individual words in the delayed word recall such as hotel or river or individual items in the CES-D depression scale, e.g., on how many days during the past week did you feel sad.  We decided on the two primary outcomes (CES-D depression scale at age 65 and a cognition measure combining delayed word recall and letter fluency at age 65) because these are important and well understood measures of mental health and cognition, and we thought these outcomes would be more sensitive than others (i.e., have more power) if playing football had an effect.  Other outcomes were considered as secondary outcomes.

A different approach than choosing just two primary outcomes would have been to consider many primary outcomes (e.g., all the cognition and mental health tests at each of the three ages, 33 in total) and correct for multiple testing.  Examples of methods that control for the familywise multiple testing error rate are Bonferroni-Holm \cite{holm1979simple} or splitting in which the data is randomly divided into a planning sample and an analysis sample, hypotheses are chosen to test based on examining the planning sample and then the chosen hypotheses are tested on the analysis sample using Bonferroni-Holm with the number of hypotheses now being the number of chosen hypotheses from the planning sample \cite{cox1975note,rubin2006method,heller2009split}.\footnote{These methods control the familywise error rate, another alternative is to control the false discovery rate through a method such as Benjamini-Hochberg \cite{benjamini1995controlling}; see \cite{benjamini2020selective} for a good general review of methods for controlling for multiple testing.}  We compare the following three methods in a simulation study: (i) choose one outcome to test a priori; (ii) test all outcomes using Bonferroni, i.e., for a familywise Type I error rate of $0.05$, an outcome needs to have a $p$-value less than $\frac{\mbox{.05}}{\mbox{\# of outcomes}}$ for the null hypothesis of no treatment effect on it to be rejected and (iii) splitting, where the planning sample is $1/3$ of the sample and we choose the outcome with the biggest association with the treatment on the planning sample as measured by the Wilcoxon signed rank statistic  and test that outcome on the analysis sample.  We consider a setting of 100 outcomes, where the treatment has an effect on one outcome but not on the other 99 outcomes; 500 matched pairs; the matched pair differences in outcomes are independent standard normals; and the Wilcoxon signed rank test is used.  We assume that the method that chooses one outcome to test a priori has a $2/3$ chance of choosing the outcome that the treatment affects, so there is good but not perfect a priori knowledge about which outcome is affected.  We first compare the three approaches on randomized trials and then on observational studies.

\begin{figure}
\includegraphics{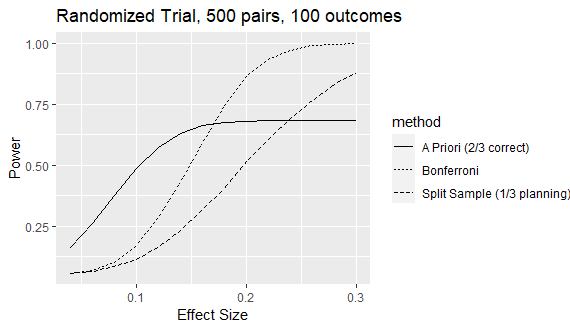}
\caption{Power for detecting effect in first outcome in matched pair randomized study with 500 pairs.  There are 100 outcomes.  The treatment's effect size on the first outcome is on the $x$ axis.  The treatment has no effect on the remaining 99 outcomes.}
\label{power.randomized.trial.fig}
\end{figure}

Figure \ref{power.randomized.trial.fig}  shows the power for different effect sizes for a randomized trial setting where we are not worried about unmeasured confounding.  For a smaller effect size, choosing an outcome a priori is better than Bonferroni but Bonferroni is better for a larger effect size.  Bonferroni always beat splitting, which is consistent with Cox's results in \cite{cox1975note}.

Now consider observational studies.  Unlike in randomized trials, bias from unmeasured confounding is a concern.  In an observational study, evidence for a causal effect is typically only compelling if it is insensitive to some bias.   Susan Ellenberg, former president of the Society for Clinical Trials, said \cite{ellenberg2018}:
\begin{quote}
Relative risks of less than 2 (or maybe even 3 or 4) found in observational studies should be viewed skeptically, no matter how many zeroes follow the decimal point.
\end{quote}
Robert Temple, former director of drug evaluation at the Food and Drug Administration, said \cite{taubes1995epidemiology}:
\begin{quote}
My basic rule of thumb is if the relative risk isn’t three or four, forget about it.
\end{quote}
Marcia Angell, former editor of the {\it{New England Journal of Medicine}}, said \cite{taubes1995epidemiology}:
\begin{quote}
As a general rule of thumb, we are looking for a relative risk of three or more [before accepting a paper for publication], particularly if it is biologically implausible or if it's a brand-new finding.
\end{quote}
Sir Richard Doll suggested that no single epidemiologic study is persuasive by itself unless the lower limit of its 95\% confidence level falls above a threefold increased risk \cite{taubes1995epidemiology}.  Doll's suggestion can be understood in terms of Rosenbaum's $\Gamma$ sensitivity analysis (see \ref{sec: design}) as saying that a study is not persuasive unless it is sensitive (maximum $p$-value $<0.05$) to a bias of $\Gamma =3$ \footnote{Specifically, for a stratified study with a binary outcome (so that the data consist of a $2\times 2\times S$ contingency table where there are $S$ strata), then a study is insensitive to bias at $\Gamma =3$ at level $0.05$ if the lower 95\% confidence interval of the odds ratio for the treatment effect has limit $>3$ \cite{rosenbaum1995quantiles}.}.  Then, according to Doll's suggestion, if there is a treatment effect of the magnitude we hope for, then we would like to be able to say the study is insensitive to bias at $\Gamma =3$.  The probability of being able to make this statement when there is a treatment effect of a given magnitude is the power of sensitivity analysis for that treatment effect at $\Gamma =3$ \cite{rosenbaum2010design}.  \footnote{The power is considered under the ``favorable'' situation that the treatment has an effect and there is in fact no unmeasured confounding bias, see \cite{hansen2014clustered}, \S 3.3. Even though there is in fact no unmeasured confounding bias in this situation, we would typically not be able to know this and so would like to have power that allows for a bias up to $\Gamma =3$ {\textendash} see the quotes above from Ellenberg, Temple, Angell and Doll.  If we cannot know when we are in the favorable situation, and if we may not be, then why should we be interested in the power computed in the favorable situation?  Hansen, Rosenbaum and Small \cite{hansen2014clustered} explained, ``In computing power in the favorable situation we are asking about the ability of a particular research design and method of analysis to discriminate between two situations in which we know unambiguously what answer is desired of the sensitivity analysis. If there is a moderate bias in treatment assignment and no treatment effect, then we hope that the sensitivity analysis will tell us that the observed association between treatment and outcome can be explained by a bias of magnitude $\Gamma$, and by construction [of the sensitivity analysis] we take only a risk of at most $\alpha$ [the specified Type I error rate] that the sensitivity analysis will report otherwise in this situation. If there is no bias in treatment assignment, $\Gamma = 1$, and there is a treatment effect, then we hope to reject the null hypothesis $H_0$ of no effect, and the power of a sensitivity analysis in the favorable situation is the chance that our hope will be realized.  If there were both a bias in treatment assignment and also a treatment effect, then we must be ambivalent about rejecting the hypothesis of no effect, $H_0$, even though it is false. Suppose, for example, that there was a large bias in treatment assignment and a small treatment effect, so that rejection of $H_0$ is nearly assured for all small or moderate $\Gamma$, then, we cannot be pleased to reject $H_0$ for small or moderate $\Gamma$ because we know we would also have rejected $H_0$ in this situation had it been true.'' \label{footnote_favorable}}
The magnitude of bias $\Gamma$ that we are concerned about and want to have high power for in an observational study will not always be $\Gamma =3$ but it will typically be some $\Gamma >1$.\footnote{Good design of an observational can reduce the $\Gamma$ (the magnitude of unmeasured confounding) that we are concerned about. See for example the discussion in Section \ref{sec: design} of \cite{lehman1987long}'s study of the effect of the loss of spouse or child in a car crash on long term mental health \cite{lehman1987long}.}  Rosenbaum writes \cite{rosenbaum2005sensitivity}:
\begin{quote}
[In an observational study] there is usually the concern that some important baseline differences were not measured so that individuals who appear comparable may not be.
\end{quote}

\subsubsection{Comparison of Methods in Observational Studies}

\begin{figure}
\includegraphics{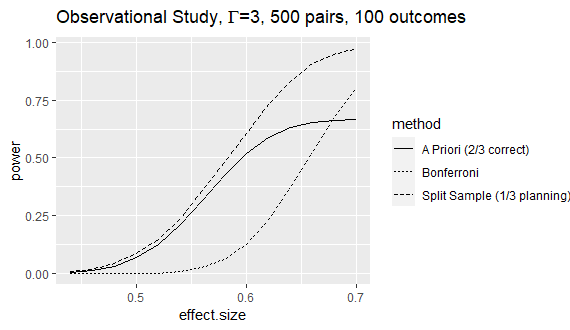}
\caption{Power of sensitivity analysis allowing for unmeasured confounding up to $\Gamma =3$ for detecting an effect in the first outcome in a matched pair observational study with 500 pairs.  There are 100 outcomes.  The treatment's effect on the first outcome is the effect size on the $x$ axis.  The treatment has no effect on the remaining 99 outcomes.}
\label{power.obs.study.fig}
\end{figure}

In Figure \ref{power.obs.study.fig}. we consider the power of sensitivity analysis for an observational study at $\Gamma =3$ for the same methods and setting as for the randomized trial in Figure \ref{power.randomized.trial.fig} except with lager effects sizes (needed to have any power at $\Gamma =3$ compared to  $\Gamma =1$).  In the observational study setting in Figure \ref{power.obs.study.fig}, splitting is best for all effect sizes, beating both choosing one outcome a priori with $\frac{2}{3}$ chance of choosing the correct outcome (out of 100 outcomes) and Bonferroni.\footnote{Section 5 of the Supplemental Materials (see in particular Figure 4 Supplemental Materials) considers an alternative method of analysis for an observational study {\textendash} linear regression with the extended omitted variable bias sensitivity analysis of \cite{cinelli2020making} {\textendash} and shows qualitatively similar results as Figure \ref{power.obs.study.fig}.}  In contrast, in the randomized trial setting of Figure \ref{power.randomized.trial.fig}, splitting is never best.

\begin{figure}
\includegraphics{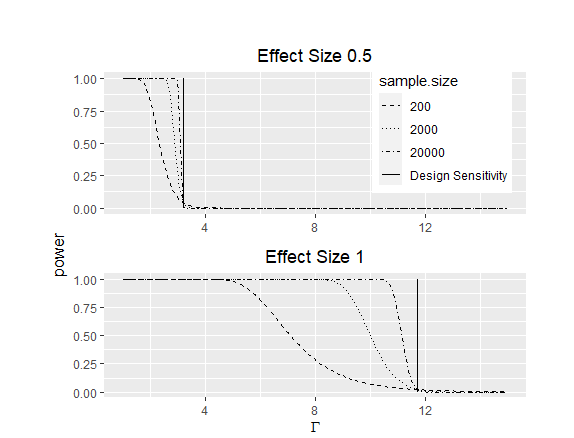}
\caption{Approximate power of a sensitivity analysis for Wilcoxon's signed rank test statistic when there is no bias from unmeasured covariates, the treatment has an additive constant effect and the treated-minus-control differences in observed responses are independently sampled from a normal distribution for samples of size 200, 2000 and 20,000 pairs.  The solid vertical line is the design sensitivity.  The power tends to 1 as the number of pairs increases for $\Gamma$ less than the design sensitivity and 0 for $\Gamma$ greater than the design sensitivity.  The power is computed for a $0.05$ level test.}
\label{fig:effectsize}
\end{figure}

Let's consider some intuition for why splitting is best in the observational study setting of Figure \ref{power.obs.study.fig}.  Consider the comparison of splitting to a priori choosing an outcome we think is best.  The disadvantage of splitting is that it loses $\frac{1}{3}$ of the sample as only the $\frac{2}{3}$ analysis sample is used to test the hypothesis of no effect.  But in an observational study, an increase in sample size has only a limited effect on power for $\Gamma\gg 1$. Consider the top panel of Figure \ref{fig:effectsize} that shows the power allowing for various amounts of unmeasured confounding $\Gamma$ for a matched pair study using Wilcoxon's signed rank statistic with effect size $0.5$ for sample sizes of $200$, $2000$ and $20,000$ pairs.
\footnote{Sensitivity analysis for Wilcoxon's signed rank test can be done in \texttt{R} using the \texttt{senWilcox} function in the \texttt{DOS2} package}   For most $\Gamma$, there is only a small increase in power going from a sample of 200 to 2000 pairs and only a slight increase in power going from 2000 to 20,000 pairs.  The solid vertical line is at the design sensitivity where for $\Gamma$ less than the design sensitivity, the power tends to one as the sample size increases to infinity and for $\Gamma$ larger than the design sensitivity, the power tends to 0 \cite{rosenbaum2004design}.  Here, the design sensitivity is $\Gamma =3.17$.  The bottom panel of Figure \ref{fig:effectsize} shows the power for effect size 1.  Similar to the top panel, there is only a small increase in power as the sample size increases from 200 to 20,000 for most $\Gamma$.  But now the power is much higher for a range of $\Gamma$, e.g., for $\Gamma =5$, the power is close to 0 for effect size $0.5$ for all sample sizes but the power is close to 1 for effect size $1$ for all sample sizes.  The design sensitivity increases from 3.17 for effect size $0.5$ to 11.72 for effect size $1$.  The advantage of splitting is that we may be able to figure out from the $\frac{1}{3}$ planning sample which outcome(s) have higher design sensitivities and focus on testing them in the analysis sample.  In the setting of Figure \ref{power.obs.study.fig} with 500 pairs and 100 outcomes, the probability of correctly selecting the one outcome that the treatment has an effect on from the $\frac{1}{3}$ planning sample is close to 1.\footnote{See \cite{heller2009split}, \S 3.3, for formula for the probability of correct selection.}  As seen in Figure \ref{fig:effectsize}, there can be a huge gain in power for choosing an outcome with higher design sensitivity and the loss in power from only being able to use the $\frac{2}{3}$ analysis sample rather than the full sample is small.



\begin{figure}
\includegraphics{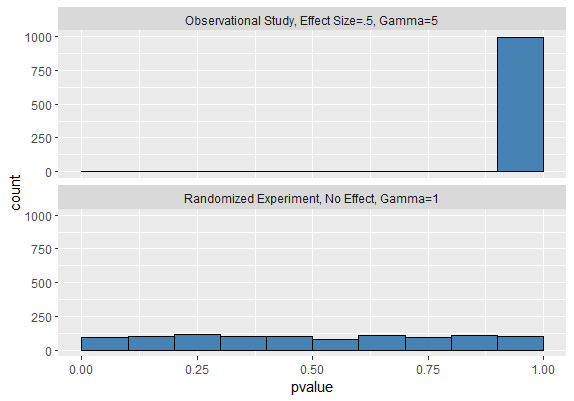}
\caption{Histogram of $p$-values for 1000 simulations of 500 matched pairs with differences that are normally distributed with variance 1 in which (a) there is a treatment effect of size 0.5 and the $p$-values are for a sensitivity analysis for the Wilcoxon signed rank test that allows for unmeasured confounding $\Gamma =5$ and 
(b) there is no treatment effect and the $p$-values are for the Wilcoxon signed rank test that assumes no unmeasured confounding. The design sensitivity for (a) is 3.17 so as the number of matched pairs increases to infinity, the $p$-values converge to 1.}
\label{pvalues_histogram}
\end{figure}

Next, let's consider intuition for the Bonferroni vs. splitting comparison.  Bonferroni dominated for randomized trials in Figure \ref{power.randomized.trial.fig} but splitting dominated for observational studies in Figure \ref{power.obs.study.fig}.  Bonferroni permits all hypotheses to have $p$-values that are uniformly distributed so it pays a large price for testing many hypotheses.  In randomized trials in which we want to test the hypotheses for $\Gamma =1$ (no unmeasured confounding), a $p$-values will in fact be uniformly distributed if the null is true so this price needs to be paid.  But in observational studies in which we want to test the hypotheses for some $\Gamma >1$ to allow for unmeasured confounding, most $p$-values for true nulls are stochastically much larger than the uniform distribution {\textendash} see Figure \ref{pvalues_histogram} for an illustration. Bonferroni is then conservative but splitting can screen out those hypotheses that never had a chance, greatly reducing the extent of the multiplicity correction \cite{zhao2018cross}.\footnote{\cite{zhao2018cross} consider an alternative to splitting, cross-screening,which splits the data in half at random, uses the first half to plan a study carried out on the second half, then uses the second half to plan a study carried out on the first half, and reports the more favorable conclusions of the two studies, correcting using the Bonferroni inequality for having done two studies.  Splitting is better than cross-screening as the sample size grows because the planning sample only needs to be large enough to make sensible plans. But for small samples, the planning sample may be too small to make sensible plans and cross-screening has better power \cite{zhao2018cross}.}

An additional advantage of splitting is that it permits exploratory data analysis to be done freely on the planning sample, which may generate unanticipated insights and permits human judgment to play an informed role between exploratory analysis of the planning sample and focused confirmatory analysis of the analysis sample.  Algorithmic procedures that address multiple testing, such as Bonferroni, do not leave a role for judgment; rather, their form must be prespecified \cite{cox1975note,heller2009split}.  Next, we illustrate these advantages of splitting in an empirical study.   

\subsubsection{Splitting the Sample in an Observational Study of the Effect of Obstetric Unit Closures}
\label{splitting.obstetric}

Between 1997 and 2003, 9 of 19 hospitals in Philadelphia closed their obstetrics units.\footnote{Four additional hospitals closed their obstetric units between 2004 and 2019 but the data we had available at the time we carried out the study in \cite{zhang2011using} were only from 1995 until 2003.}  The hospitals that closed their obstetric units were mostly community hospitals and the consequence of the closures was that the city’s large, regional teaching hospitals (whose obstetrics units remained open) were unexpectedly more crowded. Nothing comparable happened in other United States cities, where there were only sporadic changes in the availability of obstetrics units. In \cite{zhang2011using}, we considered what impact the closures in Philadelphia had on mothers and their newborns by comparing Philadelphia before and after the closures to a ``control'' Philadelphia constructed from elsewhere in Pennsylvania, California and Missouri (the three states for which we had data) over the period 1995-2003.  We constructed the control Philadelphia by matching births in Philadelphia to births in the rest of Pennsylvania, California and Missouri for 59 covariates that included year of birth and characteristics of the baby, the mother and the mother's neighborhood.  There were 132,786 births in Philadelphia and 5,998,111 control births from which we formed 132,786 matched pairs.  Because we had a large number of matched pairs, we decided to use a 10\% planning sample.

\begin{figure}
\includegraphics[scale=.33]{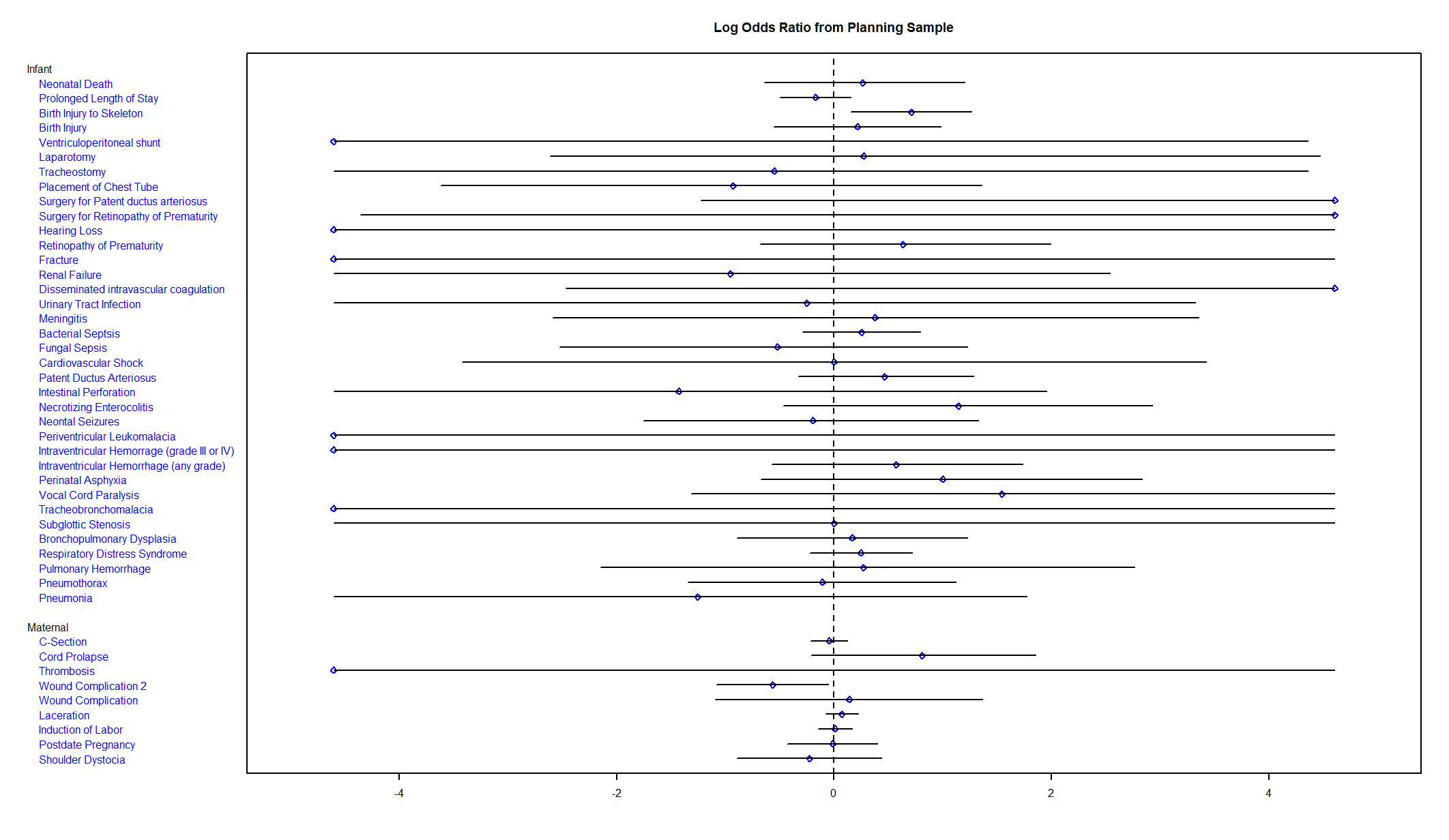}
\caption{Estimated log odds ratios and 95\% confidence intervals based on the planning sample for the effect of obstetric unit closures on various maternal and neonatal outcomes.  The intervals are left and right censored at $\log 1/100 =-4.61$ and $\log 100 =4.61$ respectively.}
\label{fig:obstetric.closure.outcomes}
\end{figure}

We did an extensive exploratory analysis of the planning sample.  We examined 45 maternal and neonatal outcomes.  Several unexpected insights emerged.  Although we knew that beginning in 2000, the city of Philadelphia intervened to slow down and organize closures, we had thought this might be a symbolic gesture. Instead, the planning sample suggested that most of the effect of the closures occurred in 1997-1999 (when five hospitals abruptly closed their obstetric units) and that after the city's intervention in 2000, there was no discernable effect of the closures.  Another unexpected insight was that before looking at any data, we thought that overcrowding in an obstetrics ward might result in an increase in C-sections and birth injuries of various kinds, but the planning sample strongly suggested a focus on birth injuries to the skeleton (which are more serious than typical birth injuries) and not a focus on C-sections.  Figure \ref{fig:obstetric.closure.outcomes} shows point estimates and 95\% confidence intervals based on the planning sample for the log odds ratio of the effect of the closures in 1997-1999 on each of the 45 outcomes.  The only outcome that has a confidence interval that does not contain 0 (i.e., does not contain an odds ratio of 1) is birth injury to the skeleton with a confidence interval of (0.16, 1.27) (i.e., (1.17, 3.55) for the odds ratio).\footnote{See \cite{zhang2011using}, \S 3.2, for details on the inference, a difference-in-difference inference for a binary outcome.  \cite{zhang2011using} also discuss the use of evidence factors and a test for bias from unmeasured confounders; Figure \ref{fig:obstetric.closure.outcomes} only concerns the first evidence factor, the comparison of the years affected by hospital closures (1997-1999) to the base years (1995-1996).  Also, the results presented here differ slightly from those in \cite{zhang2011using} as we used the recently developed \texttt{bigmatch} R package to do the matching \cite{yu2020matching}}  The planning sample suggested that several outcomes were too rare to have power to study even with the much larger analysis sample, e.g., the following two outcomes occurred only once each in the planning sample: periventricular leukomalacia (injury to the white matter around the fluid-filled ventricles of the brain) and tracheobronchomalacia (tissue that makes up the windpipe is soft and weak).  We had a meeting of the statisticians and clinicians in the study to discuss the results from the planning sample. The plan for analyzing the analysis sample that emerged from that meeting reflected results from the planning sample combined with clinical and statistical judgement.  We decided that we would focus on birth injuries to the skeleton and on whether the closures in 1997-1999 had an effect on these injuries.  If this focus had come about after examining many outcomes and various comparisons for those outcomes using the complete data, then there would naturally be concern that any significant results on the analysis sample could be the result of a fishing expedition.  However, this analytic focus came about only from looking at the planning sample, so evaluating it on the analysis sample provided a fresh test.  Using the analysis sample, we estimated that the closures increased the odds of a birth injury to the skeleton 1.67 times with $p$-value $3\times 10^{-8}$ and 95\% confidence interval (1.40, 1.99); see Figure 4 in the Supplemental Materials.
This significant effect would be sensitive to a moderate but not very small bias from unmeasured confounders of $\Gamma =1.22$. \footnote{See \cite{zhang2011using}, \S 4, for how to do sensitivity analysis for a difference-in-difference analysis with a binary outcome}

The use of splitting where we test hypotheses formulated with a planning sample on a fresh analysis sample properly controls for having selected hypotheses to test from the data (as long as if multiple hypotheses are selected from the planning sample, the multiple testing is properly controlled for in the analysis sample \cite{benjamini2020selective}).  \cite{benjamini2020selective} reviews other methods that properly control for having selected hypotheses to test from the data.  However, splitting and all these other methods do not account for two other types of selection, selective bias and selective interpretation of importance, which we will now discuss.

\subsubsection{Selective Bias}
\label{selective_bias}

\begin{figure}
\includegraphics[scale=0.6]{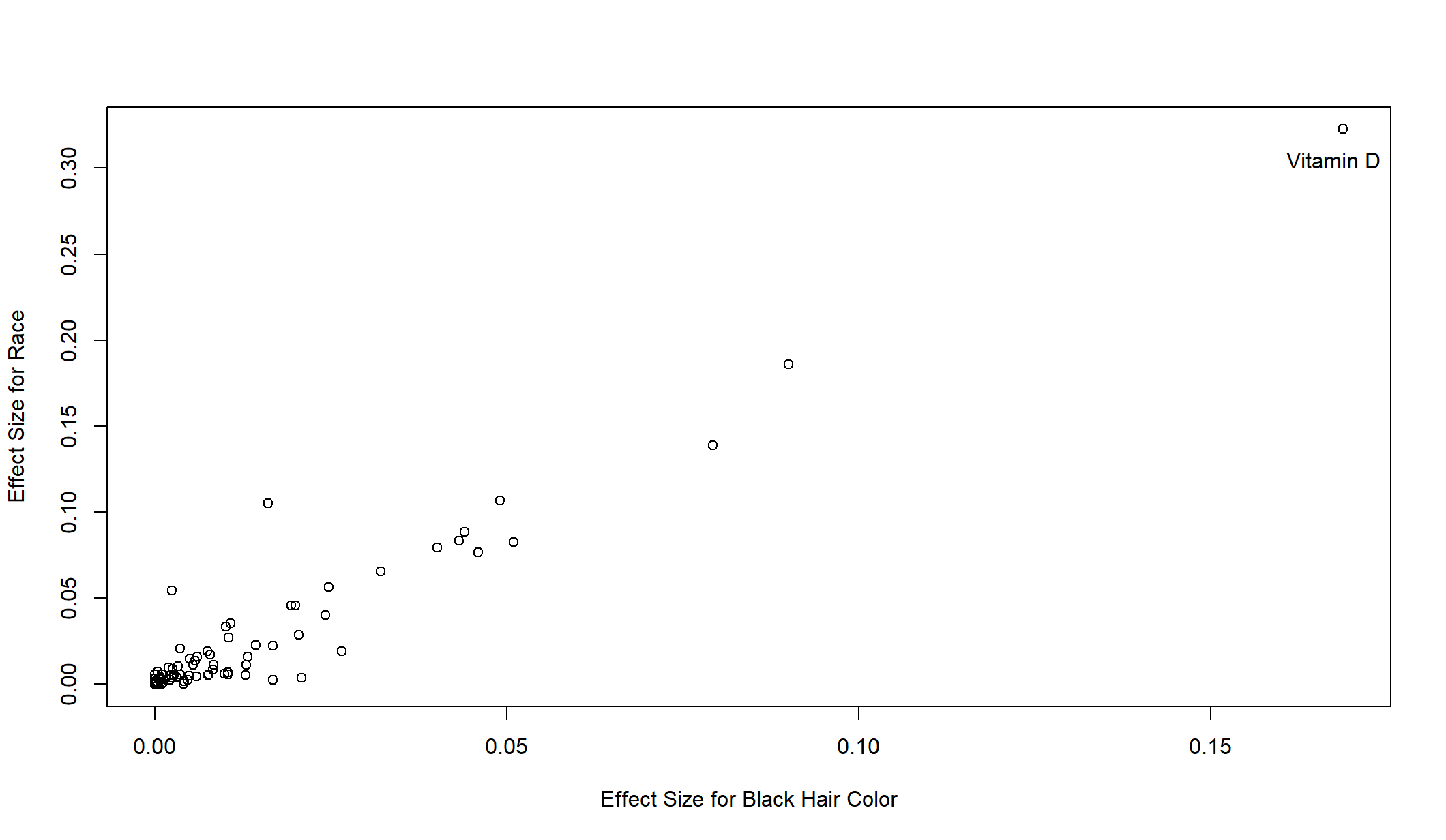}
\caption{Plot of the effect size for black hair color vs. the effect size for race for various laboratory measurement outcomes, where both effect sizes are estimated after controlling for age and gender.  The point that represents the Vitamin D outcome is labeled}
\label{fig:haircolor.race.plot}
\end{figure}

The bias from unmeasured confounding might differ between outcomes.  In splitting, when we select the outcome with the biggest association(s) with the outcome on the planning sample, we might just be selecting the outcomes with the biggest bias(es) from unmeasured confounding.  As an example, consider estimating the effect of black hair color on laboratory health measurements (e.g., cholesterol, lead and vitamins) controlling for the potential confounders age and gender.  We use data from the National Health and Nutrition Examination Survey (NHANES) from 2001-2002 on subjects 20-59 (the age range for which natural hair color at age 18 was asked) whose race/ethnicity is either non-Hispanic white or non-Hispanic black, and we consider the 75 laboratory measurements which were measured on all subjects who did not meet exclusion criteria (data in supplemental materials).  Presumably black hair color has no casual effects but there are unmeasured potential confounders, e.g., race.  70\% of blacks have black hair color compared to only 5\% of whites.  Figure \ref{fig:haircolor.race.plot} is a scatterplot for the laboratory measurements of the effect size for black hair color vs. the effect size for race (where the effect sizes were estimated by linear regression and the R package \texttt{effectsize}).  The laboratory measurement with the biggest effect size for black hair color is Vitamin D (effect size $=.17$), and it also has the biggest effect size for race (effect size $=.33$).  A major source of Vitamin D for people is that when skin is exposed to sunlight, the sun’s ultraviolet B (UVB) rays hit cholesterol in the skin cells, providing the energy for vitamin D synthesis to occur.  People with dark skin get less Vitamin D from the sun because the pigment (melanin) in dark skin does not allow the skin to absorb as much UV radiation.  Thus, race is a confounder for the effect of black hair color on Vitamin D.  After controlling for race, there is no evidence for an effect of hair color on Vitamin D {\textendash} the estimated effect size is $0.00$ with a 95\% confidence interval of $(0.00,0.01)$.

\subsubsection{Selective Interpretation of Importance}

Consider a hypothetical study of the effect of playing football on 26 neuroimaging measures $A$, $B$, ..., $Z$.  Consider a researcher who would a priori have ranked the importance of the measures for brain health as follows: 1. $C$; 2. $G$; 3. $Q$; ...; 25. $L$; 26. $B$.  But suppose that the researcher did not write down any a priori ranking and used splitting to select measure(s) from the planning sample to study in the analysis sample.  Also suppose the researcher has a strong a priori belief that playing football has an important harmful effect on the brain.
Now suppose measure $L$ is the only measure that appears to be significantly affected by playing football from the planning sample, and the analysis sample also provides evidence that playing football affects $L$.  Even though the researcher would have considered measure $L$ the second least important measure had the researcher ranked the measures a priori, the researcher might now think of reasons why $L$ is very important.  There has been a selective interpretation of importance.  This would not necessarily be because the researcher was being deceptive, but might instead be the researcher's heartfelt interpretation that has been colored by confirmation bias.  The philosopher Francis Bacon wrote \cite{bacon1620}:
\begin{quote}
The human understanding when it has once adopted an opinion ... draws all things else to support and agree with it. And though there be a greater number and weight of instances to be found on the other side, yet these it either neglects or despises, or else by some distinction sets aside or rejects[.]
\end{quote}
Note that selective interpretation of importance might arise not just when splitting is used but anytime a study considers multiple outcomes and does not a priori specify the importance of the different outcomes, e.g, if the outcomes reported are those that are significant when the false discovery rate is controlled.  

We consider three possible remedies for selective interpretation of importance.  One is to a priori rank the importance of all the outcomes.  However, this is hard to do when there are many outcomes.  Another remedy is to a priori decide on the most important outcome(s) and only test these more important outcomes(s) in the primary analysis.  A primary analysis could be accompanied by secondary analyses, and findings from secondary analyses could be followed up in future studies rather than regarded as confirmatory.  However, some observational studies focus on events or data that are unique or not easily replicable anytime soon such as the study of the obstetric unit closures in Philadelphia.  For such studies, splitting offers a way to make confirmatory findings for a variety of possible outcomes but carries the risk of selective interpretation of importance.  A possible remedy for selective interpretation of importance when using splitting is to a priori specify an outcome(s) that captures the broad effect of interest that will automatically be included in the analysis sample while allowing for selecting other outcomes from the planning sample.  For example, in the obstetric unit closure study, we might a priori specify that infant death, whether there is at least one infant complication ($\geq 1$ infant complication) and whether there is at least one maternal complication ($\geq 1$ maternal complication) will be included in the analysis sample since these capture the most important adverse infant effect, a broad measure of adverse infant effects and a broad measure of adverse maternal effects respectively.  Then, after we examine the planning sample and selected birth injury to the skeleton as an additional outcome to test, we would test four outcomes in the analysis sample and in order to control the familywise error rate at .05, we would form $(1-.05/4)\times 100 = 98.75\%$ confidence intervals on the analysis sample.  These confidence intervals are shown in Table \ref{obstetric.analyis.outcomes}.  For maternal complications, there appears to be little effect.  For infant complications, while the effect for the closures on $\geq 1$ infant complication is not as large as for the specific infant complication of birth injury to skeleton, it is still clinically significant (estimated odds ratio of 1.31) and statistically significant.  For infant death, the effect of the closures on any infant death is not statistically significant but the point estimate is that there is an adverse effect.  Overall, this supports the finding from the outcome of birth injury to skeleton that there is an adverse effect of the closures on infant outcomes.  Suppose counterfactually that there had been an effect on birth injury to the skeleton but no significant effect on any of infant death, $\geq 1$ infant complication or $\geq 1$ maternal complication, and it was claimed post hoc that because birth injury to the skeleton is a particularly important outcome, obstetric units have an important effect on infant outcomes.  The findings of no significant effect on the three a priori defined important outcomes would perhaps temper this post hoc claim. 

\begin{table}[ht!]
\caption{Effect of Obstetric Unit Closures in Philadelphia: Confidence intervals for the odds ratio for the effect of obstetric unit closures on the analysis sample for one outcome selected from the planning sample (birth injury to skeleton) and three outcomes chosen a priori ($\geq 1$ maternal complication, $\geq 1$ infant complication and infant death).}
\begin{center}
\begin{tabular}{|c|c|c|}
\hline
Outcome & Odds Ratio & 98.75\% Confidence Interval \\ \hline
Birth injury to skeleton & 1.67 & (1.33, 2.09) \\  \hhline{|=|=|=|}
$\geq 1$ maternal complication & 1.03 & (0.99, 1.08) \\
$\geq 1$ infant complication & 1.31 & (1.19, 1.43) \\
Infant death & 1.12 & (0.81, 1.56) \\ \hline
 \hline
\end{tabular}
\end{center}
\label{obstetric.analyis.outcomes}
\end{table}

\subsubsection{Discussion: Multiple Outcomes in Observational Studies vs. Randomized Trials}
\label{sec: discussion: multiple outcomes}

In a well designed randomized trial, uncertainty about treatment effects comes mainly from sampling variation {\textendash} unobserved bias from unmeasured confounding is avoided by randomization.  But in an observational study, unobserved bias from unmeasured confounding can contribute substantially to uncertainty.  Generally, an increase in sample size in an observational study has only a limited effect on this type of uncertainty \cite{rosenbaum2004design}.  In contrast, splitting {\textendash} that is, discarding a small part of the sample size to improve the
study design {\textendash} holds the realistic prospect of making the study less sensitive to unobserved biases, so less uncertain in a sense \cite{heller2009split}.  In a moderately large observational study, the sample size is being wasted if it is used only to reduce sampling variability, and not used to improve design, because reduced sampling variability has only a very limited impact on bias from unmeasured covariates, a key source of uncertainty.

In randomized trials, it is often suggested that one primary outcome be chosen \cite{stanley2007design,chan2013spirit}.  The considerations in the above paragraph suggest there is more of an incentive to consider multiple primary outcomes in a moderately large observational study than in a randomized trial.  However, some of the same considerations which suggest restricting to one primary outcome in a randomized trial also apply to an observational study,  including the following \cite{tukey1980we}:
\begin{itemize}
\item {\it{Power}}.  Power when looking at multiple outcomes can still be a concern when the sample size is not huge.  The following advice from Tukey \cite{tukey1980we} for randomized trials also has some applicability to observational studies:
\begin{quotation}

Problems of multiplicity have been too little recognized. To say in advance that we will look at one of 12 analyses is to give many hostages to fortune. If the results of the 12 analyses are statistically independent, at least one will be ``significant at 5 percent'' a large fraction, $1 - .95^{12} = .46$, of the time. This is ordinarily an unacceptable Type I error. If we protect against this by going to 5\%/12 as our significance level, then, IF the results are highly correlated, we shall have been highly wasteful, concluding less about our question than we should.

I see no real alternative, in most truly confirmatory studies, to having a single main question-in which a question is specified by ALL of design, collection, monitoring, AND ANALYSIS.

It may be wise, sometimes, to include alternative analyses, but we ought to regard any such case as a failure of statistical theory. To calculate two statistics and take the larger becomes a single analysis as soon as we know enough about the null distribution of ``the larger of this and of that''! If neither statistical theory or computer rerandomization can tell us enough about the actual null distribution, how can we be content?
\end{quotation}
\item {\it{Selective interpretation}}.  Having one primary outcome avoids selective interpretation in interpreting the primary analysis \cite{chan2013spirit}.  Having say two primary outcomes leaves open the possibility that when an investigator has an a priori positive view of the treatment and the results are inconsistent between the two outcomes, the results may be more likely to be interpreted as favorable to the treatment regardless of which of the two outcomes was more in favor of the treatment.
\item {\it{Clarity}}.  Having one primary outcome makes the study clearer to understand for a public audience.
\item {\it{Transparency}}.  Related to clarity and avoiding selective interpretation, having one primary outcome aids transparency (making evidence evident).  Transparency encourages critical discussion \cite[Ch. 6]{rosenbaum2010design}. 
 Good critical discussion is important for advancing our understanding of what treatments are effective.  John Stuart Mill said \cite{mill1859liberty},
    \begin{quote}
    The beliefs which we have most warrant for have no safeguard to rest on, but a standing invitation to prove them unfounded.
    \end{quote}
    Karl Popper said \cite{popper1972objective},
    \begin{quote}
    The objectivity of all science...is linked with its criticizability.
    \end{quote}
\end{itemize}

Having one primary analysis does not preclude secondary and exploratory analyses; it just distinguishes such analyses \cite{rosenbaum2020modern}.  Also, having one primary analysis along with additional secondary and exploratory analyses does not preclude interpreting the totality of evidence \cite{pocock2016primary}.  However, prespecification of secondary analyses in a protocol and accounting for multiplicity are helpful for properly interpreting the totality of evidence.  We now consider these issues in the context of how well the evidence corroborates an elaborate theory.  

\paragraph{Elaborate Theories}
\label{elaborate.theories}

An elaborate theory is a set of multiple predictions about how the data would look if the treatment has an effect.  In his seminal paper on observational studies \cite{cochran1965planning}, Cochran advocated using elaborate theories, attributing the advice to Fisher:
\begin{quote}
When asked in a meeting what can be done in observational studies to clarify the step from association to causation, Sir Ronald Fisher replied: ``Make your theories elaborate.''  The reply puzzled me at first, since by Occam's razor the advice usually given is to make theories as simple as is consistent with the known data. What Sir Ronald meant, as the subsequent discussion showed, was that when constructing a causal hypothesis one should envisage as many different consequences of its truth as possible, and plan observational studies to discover whether each of these
consequences is found to hold...[T]his multiphasic attack is one of the most potent weapons in observational studies.
\end{quote}

In Fisher's advice to use elaborate theories, one needs to make the predictions about how a treatment effect would manifest before seeing the data {\textendash} there is no value in an elaborate theory built after the fact to fit unanticipated patterns in the data \cite{rosenbaum2010design}.  Thus, an elaborate theory and a plan to test it should be specified in the protocol.  In our football study, we specified in our protocol an elaborate theory that if football harms long term mental functioning and there are not unmeasured differences between the football players and controls, then 
\begin{enumerate}
\item[(i)] the football group should have worse long term mental functioning than the matched overall control group; 
\item[(ii)] the football group should have worse long term mental functioning than each of the two matched control groups, the matched non-collision sports control group and the matched non-sports control group; 
\item[(iii)] the difference between the football group and the closer of the two matched control groups on long term mental functioning should be greater than the difference between the two matched control groups; and 
\item[(iv)] since acute cognitive dysfunction following concussions or mild traumatic brain injury resolves within 3 months in most patients, there should not be differences in medium-term outcomes before 35 years of age between the football and matched control group.
\end{enumerate}
The comparison (i) between the football group and the matched overall control group might be considered the usual primary analysis.  This usual primary analysis can be synchronized with testing the elaborate theory by using testing in order \cite{rosenbaum2008testing}.  We can first test (i) and if a significant difference at the .05 level is found between the football and overall matched control groups, go on to test (ii) and if significant differences are found, go on to test (iii) and so on  \footnote{If a significant difference at the .05 level is found in (i), then each of the two tests in (ii), football vs. the matched non-collision sports players and football vs. the matched non-sports players can be done at the .05 level while still controlling the familywise Type I error rate at the .05 level because it is a sequentially exclusive partition meaning that if the null hypothesis that football and the matched overall control group are the same is false, then at most one of the null hypotheses in (ii) can be true \cite{rosenbaum2008testing}.}  The familywise type I error rate for all the hypotheses in the elaborate theory is controlled at level .05.  This ordered testing of the elaborate theory accomplishes three valuable things.  First, it results in a graded assessment of corroboration with the elaborate theory, i.e, evidence for (i)-(iv) is greater than evidence for just (i)-(iii) which is greater than evidence for just (i)-(ii).  Second, it controls the familywise Type I error rate when multiple hypotheses are tested.  Third, it prioritizes comparisons, in particular prioritizing the primary analysis.  If say the comparisons were treated equally and Bonferroni was used to control for the multiple comparisons, then as Rosenbaum \cite{rosenbaum2010design} says
\begin{quote}
the investigator might discover, to her shock and dismay, that if she had used the first control group alone, it would have had outcomes significantly different from the treated group, and if she had used the second control group alone, it too would have had outcomes significantly different from the treated group, but because she used both control groups and did two tests, neither difference is significant after correcting for multiple testing. This...is not appropriate.
\end{quote}

Although ordered testing of an elaborate theory has considerable advantages, one negative aspect is that although one might want to put a higher priority on certain aspects of the theory, one might want some information on all aspects.  A different approach to assessing the evidence for an elaborate theory is to make inferences about the fraction of the elaborate theory that is corroborated as well as individual aspects of the theory \cite{karmakar2020assessment}. Further research on ways to assess the evidence for an elaborate theory would be useful.

\subsection{Subgroups}
\label{sec: subgroups}

In the observational study protocols presented in Section \ref{sec: case studies}, the planned analysis was for the whole study population and did not consider subgroups.  For the football study, one subgroup of interest is position {\textendash} e.g., lineman, running back, quarterback, or wide receiver {\textendash} which each might be exposed to a different frequency and severity of head impacts and consequently be affected differently by playing football.  Unfortunately our data set, the WLS, does not provide information about position.  However, the WLS does provide information on other subgroups of potential interest such as parents' income, parents' education and student's other high school activities (e.g., participation in student government, participation in musical groups).  How should subgroups be incorporated into the planned analysis for an observational study protocol?

Let's first consider some traditional viewpoints from randomized trials.  Tukey distinguished between focused clinical trials and clinical inquiries\cite{tukey1977some}:
\begin{quote}
{\it{The focused clinical trial}}...is a trial in which both the class of patients and the endpoint to be considered are clearly specified in the protocol.

{\it{The clinical inquiry}}...is where some intervention or therapy is hoped to be of help to some class of patients, not specified in advance, and where, consequently, we go in for massive data collection and for analysis of results for each of many classes of patients (by age, sex, previous medical history, prognosis, and symptoms, for example). (There also may be separate analyses for different end points.)  The statistician must, I believe, call attention to the multiplicity of questions which any such inquiry poses, and he must, therefore, face up to the influence of this multiplicity on the strength of the evidence resulting from the inquiry.
\end{quote}

Because of the problem of multiplicity in clinical inquiries, Tukey asserted,
\begin{quote}
I do not believe that a clinical inquiry, by itself, is likely to be an ethically satisfactory means of providing definitive evidence that an intervention or therapy is an improvement.  To say this is not to say there should be no clinical inquiries. Quite the contrary. clinical inquiries may often play a very crucial and very useful role. However, at a time before such an inquiry has reached trustable conclusions, it will ordinarily be best to initiate, or to embody in the continuing clinical inquiry, a single focused clinical trial (or, possibly, a few such) from which one can come more rapidly to trustable conclusions.
\end{quote}


Peto and co-authors \cite{peto1995large} suggested focusing on the overall effect of the treatment in a randomized trial:
\begin{quote}
The largest and most important bias that is still very commonly introduced during the statistical analysis of randomized trials is that produced by unduly data-dependent emphasis on the results in particular subgroups. If the trial treatment is in fact largely or wholly ineffective then such subgroup analyses may well engender false positive results, while if the trial treatment is in fact moderately effective then they may well engender false negative results...Hence...it is often the overall results of a trial that should chiefly be emphasized.  \footnote{Peto was not arguing against exploring the data but that these exploratory analyses should not be considered primary analyses {\textendash} in fact he suggested in another place that one should always do subgroup analyses but never believe them \cite{ellenberg2018}.}.
\end{quote}
Peto and co-authors based their reasoning in part on the assumption 
that if the treatment is superior for some subgroups, it will not be inferior for other subgroups, i.e., that there are no qualitative interactions \cite{peto1995large}: 
\begin{quote}
If there really is, for some readily identifiable category of patients, a moderate difference between two treatments in their effects on some specific outcome then this difference might be larger or smaller in other readily identifiable categories of patient, but it is unlikely to be reversed. 
\end{quote}

Since the time Tukey and Peto wrote their articles quotead above, there have been methods developed that enable controlling the error rate when the data is used to discover subgroups \cite{lipkovich2017tutorial}.  Still, these methods are typically used for exploratory secondary analyses rather than the primary analysis of a clinical trials.  Altman and collaborators \cite{naggara2011problem}, writing in 2011, advocated,
\begin{quote}
A prudent interpretation of trial results is to limit findings that will affect clinical decisions to overall treatment effects regarding primary end points that have been carefully planned, powered, and controlled for errors. Hence subgroup findings should generally be considered as just exploratory results.
\end{quote}
Ellenberg in 2018 \cite{ellenberg2018} suggested for clinical trials:
\begin{quote}
Define a primary analysis hypothesis, with a specific analytic procedure...[Consider secondary analyses but] avoid interpreting any analysis other than the primary analysis as `definitive.'...[While there are an] ever increasing number of methods to account for multiple analyses, it is still best to rely on pre-specification of hypotheses and replication of results.
\end{quote}

\begin{figure}
\centering
\begin{subfigure}[t]{0.8\textwidth}
\centering
\includegraphics[width=\textwidth]{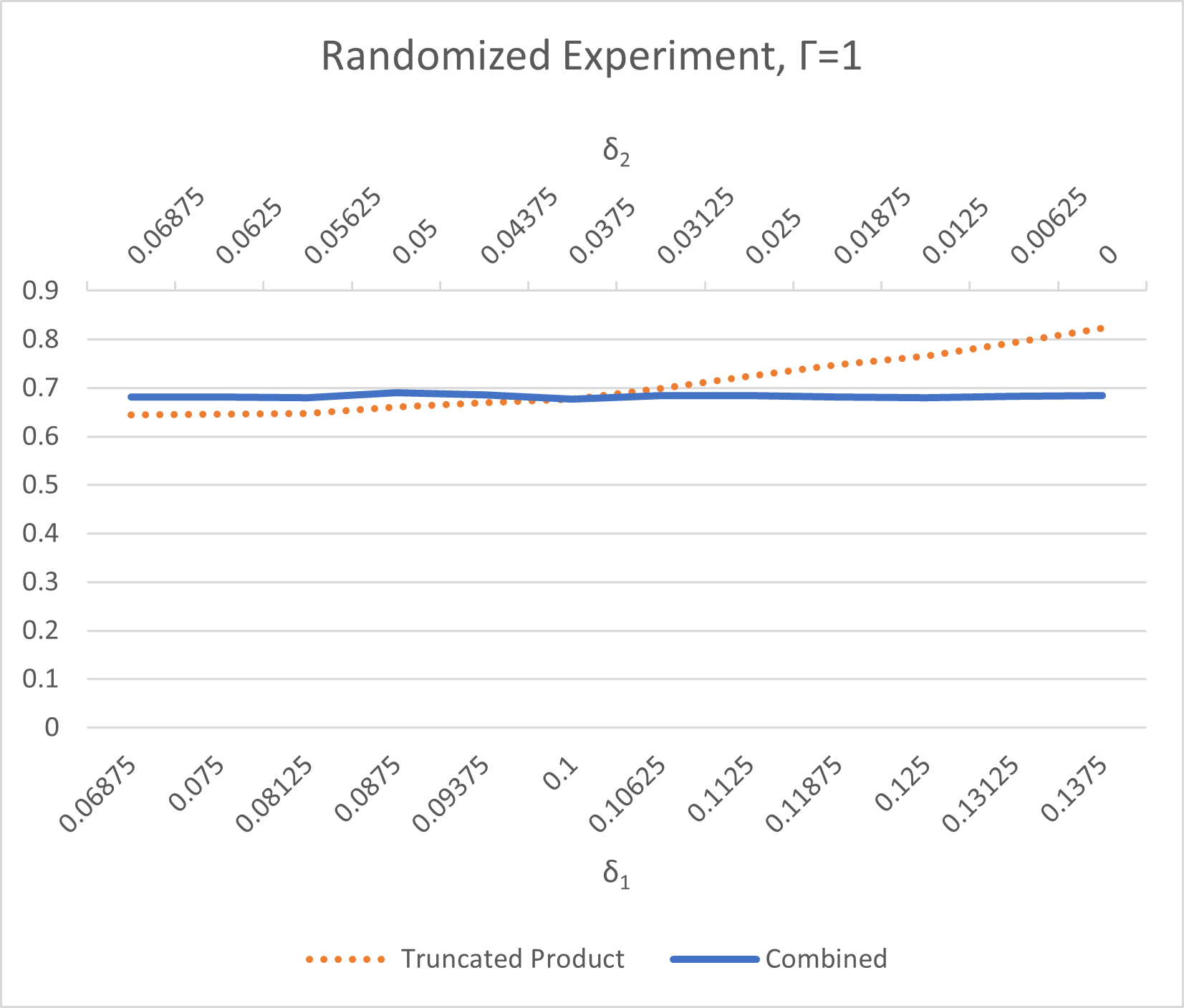}
\caption{Randomized trial settings}
\label{subgroup_power_randomized_experiment}
\end{subfigure}

\begin{subfigure}[t]{0.8\textwidth}
\centering
\includegraphics[width=\textwidth]{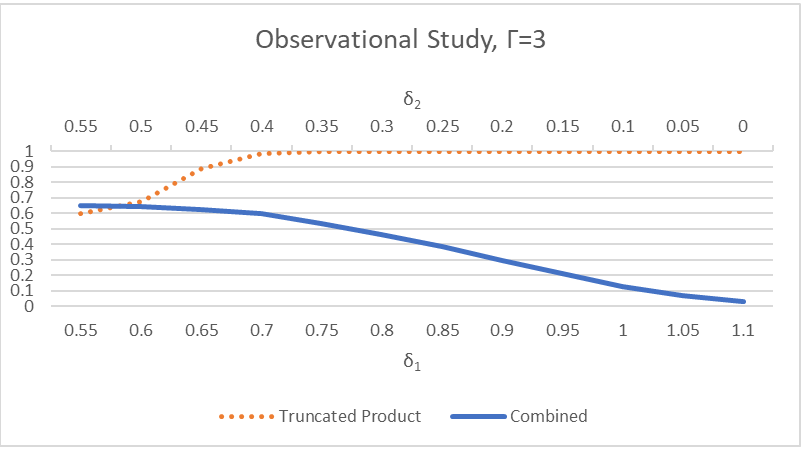}
\caption{Observational study settings}
\label{subgroup_power_obsstudy}
\end{subfigure}
\caption{Power for sampling situations in which the treatment minus control difference in a matched pair are $N(\delta_1,1)$ for the first 500 pairs in which a covariate $X$ takes on value 1 (the first subgroup) and $N(\delta_2,1)$ for the second 500 pairs in which $X$ takes on value 2 (the second subgroup).  The combined test ignores subgroups and uses all 1000 matched pairs together (using Wilcoxon's signed rank test).  The truncated product test considers the subgroups and multiplies the upper bounds under unmeasured confounding $\Gamma$ of the $p$-values from testing the first 500 pairs and from testing the second 500 pairs (using Wilcoxon's signed rank test for each group of pairs), with the $p$-values above $0.2$ replaced by 1.}
\label{subgroup_power}
\end{figure}

For incorporating subgroups into a primary analysis, should there be any different considerations for an observational study than a randomized trial?  Recall the quotes in Section \ref{sec: multiple outcomes} by Ellenberg, Temple, Angell and Doll about an observational study requiring a stronger effect than a randomized trial to provide convincing evidence of a true treatment effect because of the possibility of unmeasured confounding.  In planning an observational study, we want to have power allowing for some bias from unmeasured confounding, i.e., power for some $\Gamma >1$ in Rosenbaum's sensitivity analysis framework described in Section \ref{sec: design} where the $\Gamma$ depends on how much unmeasured confounding is of concern in the study (Doll and others suggested seeking power for $\Gamma =3$ in epidemiological studies, see Section \ref{sec: multiple outcomes}).  We now compare the incentives for incorporating subgroups into a primary analysis for a randomized trial vs. observational study under Peto et al.'s \cite{peto1995large}'s assumption that there are no qualitative interactions.  Under this assumption, in a randomized trial, if we compare the overall treated group vs. control group and find a highly significant effect, then a subgroup analysis would not change any clinical decisions.  This is the approach Altman et al. \cite{naggara2011problem} suggested:
\begin{quote}
When subgroup effects are in the opposite direction of the overall results, the most prudent approach is to consider subgroup findings as hypotheses for another trial. Until then, the best estimator of the treatment effect for any subgroup is the overall treatment effect.
\end{quote}
But in an observational study, if we find a highly significant overall treatment effect along with the (lower end of the 95\% confidence interval for the) relative risk being less than say 3, then following the suggestions of Angell, Doll, Ellenberg, Temple (discussed in Section \ref{sec: multiple outcomes}) to be skeptical of relative risks less than 3, the suggested clinical decision might be to not treat until further evidence is found.  However, if there is a subgroup where the (lower end of the 95\% confidence interval) is greater than 3, then this might be taken as evidence to treat in that subgroup and also the whole population under the no qualitative interaction assumption.  Thus, finding a subgroup which has a larger effect, even if all groups have a positive effect, has particular value in observational studies.  We present a simulation study in Figure \ref{subgroup_power}  that illustrates this point.

Figure \ref{subgroup_power} considers sampling situations in which the treatment minus control differences in outcomes in a matched pair are $N(\delta_1,1)$ for the first 500 pairs in which a covariate $X$ equals 1 (the first subgroup) and $N(\delta_2,1)$ for the second 500 pairs in which $X=2$ (the second subgroup).  Two testing methods are compared.  The combined test ignores subgroups and uses all 1000 matched pairs together (using Wilcoxon's signed rank test).  The truncated product test considers the subgroups and multiplies the upper bounds under unmeasured confounding $\Gamma$ of the $p$-values from testing the first 500 pairs and from testing the second 500 pairs (using Wilcoxon's signed rank test for each group of pairs), with the $p$-values above $0.2$ replaced by 1.\footnote{See \cite{hsu2013effect} for discussion of the truncated product test for observational studies and simulations comparing different truncation levels.  The truncated product test is implemented in the \texttt{truncatedP} function in the R package \texttt{sensitivitymult}.  Fisher's combination test that multiplies all the $p$-values is a special case of the truncated product test with truncation level $1$.  If the treatment effect is different in the different subgroups, then for some values of $\Gamma$, as the sample size increases, the $p$-values will tend to zero in some subgroups and to one in other subgroups.  If the null hypothesis of no treatment effect holds, then for $\Gamma >1$, the $p$-values will all tend 1.  Fisher's combination test is conservative when under the null hypothesis, the $p$-values have a distribution that is stochastically greater than a uniform distribution as they would if considering $\Gamma >1$ and the null hypothesis of no treatment effect holds; the truncated product with say truncation level $0.2$ is less conservative and has more power.}  Figure \ref{subgroup_power_randomized_experiment} shows power for a randomized trial setting where we test at $\Gamma =1$ since unmeasured confounding is not a concern.  We consider a range of $(\delta _1,\delta_2)$ from equal effects in the two subgroups ($\delta_1 = \delta_2 = 0.06875$) to there only being an effect in the first subgroup ($\delta_1 = .1375, \delta_2 =0)$, where the average effect $\frac{\delta_1+\delta_2}{2}$ remains the same.  The truncated product test's power increases as there is more treatment effect heterogeneity while the combined test's power remains constant.  The combined test has more power for about half the range.  Figure \ref{subgroup_power_obsstudy} shows power for an observational study setting where we allow for unmeasured confounding up to $\Gamma =3$.\footnote{The power is considered under the favorable situation that the treatment has an effect and there is in fact no unmeasured confounding bias.  See footnote \ref{footnote_favorable}}.  The effect sizes have been multiplied by 8 compared to Figure \ref{subgroup_power_randomized_experiment} as larger effect sizes are needed to have power when unmeasured confounding is allowed for.  The combined test has slightly more power when there is no treatment effect heterogeneity ($\delta_1=\delta_2 = .55$) but the truncated product test has more power for most of the range, and its advantage is dramatic for more treatment effect heterogeneity, e.g., for $\delta_1=1.1, \delta_2=0$, the truncated product test has power $1$ and the combined test has power $.03$.  Overall, in the observational study settings considered, the truncated product test is appealing if even slight treatment effect heterogeneity is a possibility while in the randomized trial settings, the truncated product test is only best if at least moderate treatment effect heterogeneity is expected.

In Figure \ref{subgroup_power}, subgroups based on only one binary covariate are considered.  If subgroups based on more covariates are considered, then the size of each subgroup may start to become small, e.g., if there were five binary covariates and $2^5=32$ subgroups were considered, the sample would be split into 32 subgroups, and power may suffer.  To address this concern, \cite{lee2018powerful} developed the submax method (implemented in the R package \texttt{submax}) that for $L$ binary covariates, splits the population into two subpopulations $L$ times, restoring the population after each split, so that successive splits do not make even smaller subpopulations, and then corrects for multiplicity using the joint distribution of these test statistics.

We now consider an example of using subgroups in an observational study about an intervention to control malaria in Africa.

\subsubsection{Example: Malaria in West Africa}
\label{garki.description}

Working with the government of Nigeria, the World Health Organization contrasted several strategies to control malaria in an observational study \cite{molineaux1980garki}.  Following \cite{hsu2013effect}, we consider the comparison of one strategy {\textendash} spraying with the insecticide propoxur together with mass administration of the drug sulfalene-pyrimethamine {\textendash} to the control of no treatment.
Assignment to treatment or control groups was based on the judgment and convenience of the investigators \cite[p.~28--30]{molineaux1980garki}. Issues that weighed on the investigators' minds in assigning treatments included practical aspects of the following:  spraying the insecticide, frequently collecting data on mosquitoes and obtaining repeated blood samples.  The outcome that we consider is a measure of the density of the malaria parasite {\it{Plasmodium falciparum}} in blood samples, specifically a post-treatment measure of the proportion of 200 microscope fields that have the parasite minus the pre-treatment proportion \cite{hsu2013effect}.  We formed 1560 matched pairs by matching treated and control subjects for age and gender \cite{hsu2013effect}.  Data from the matched pairs is provided in the supplementary materials.

\begin{figure}
\begin{center}
\includegraphics{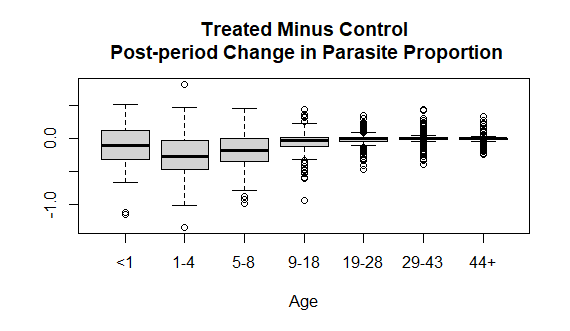}
\caption{Box plots of treated minus control difference in the post-period change (compared to the pre-period) in the proportion of microscope fields with the malaria parasite {\it{Plasmodium falciparum}} for different age groups.}
\label{agegroup.malaria.boxplot}
\end{center}
\end{figure}

In Table \ref{malaria.sensitivity}, the column ``Combined'' shows $p$-values for testing the null hypothesis of no treatment effect using these $1560$ pairs for various levels $\Gamma$ of unmeasured confounding.  There is evidence of a treatment effect up to unmeasured confounding $\Gamma =2.9$.  The data contains information on the covariates age and gender, and the treatment might have different effects for different levels of these covariates.  The original study \cite{molineaux1980garki} considered age subgroups $<1$, $1-4$, $5-8$, $9-18$, $19-28$, $29-43$ and $\geq 44$.  Figure \ref{agegroup.malaria.boxplot} shows boxplots of the treated-minus-control malaria parasite outcome by age subgroup. It looks like the treatment might have had a bigger effect in reducing malaria parasites for kids between $1$ and $8$ than for older kids and adults.  The column ``A priori age groups'' in Table \ref{malaria.sensitivity} shows $p$-values for sensitivity analyses using the truncated product test (truncation level 0.2) with the a priori chosen age subgroups from the original study.   There is evidence of a treatment effect up to unmeasured confounding $\Gamma =3.4$, so consideration of the a priori chosen age subgroups reduced sensitivity to bias compared to just combining all subjects together (sensitive at $\Gamma =2.9$).

\begin{table}[ht]
\centering\caption
{Various sensitivity analyses for the treated-minus-control difference in the malaria study in Section \ref{garki.description}.  ``Combined'' uses all 1560 pairs together.  ``A priori age groups'' uses the age groups $<1$, 1-4, 5-8, 9-18, 19-28, 29-43 and $\geq 44$ considered by \cite{molineaux1980garki}.  ``Splitting'' uses a 25\% planning sample to choose subgroups based on age and gender; the chosen subgroups are age $<9.5$ and age $\geq 9.5$. 
``Secondary'' uses the absolute differences between treated and control to choose subgroups based on age and gender; the chosen subgroups are shown in Figure \ref{absolute_tree}.  A priori age groups, Splitting and Secondary all use the truncated product to combine $p$-values with truncation point $0.2$.  All tests are Wilcoxon signed rank tests using the \texttt{senWilcox} function in the \texttt{R} package \texttt{DOS2}.  For each method, the largest p-value less than
or equal to 0.05 is in bold.}
\label{malaria.sensitivity}
\begin{tabular}{c|c|c|c|c|}
& Combined & A priori age groups & Splitting & Secondary \\ \hline
$\Gamma =1$ & 0.000 & 0.000 & 0.000 & 0.000 \\
$\Gamma =1.9$ & {\bf{0.011}} & 0.000 & 0.000 & 0.000 \\
$\Gamma =2$ & 0.072 & 0.000 & 0.000 & 0.000 \\
$\Gamma =3$ & 1.000 & 0.010 & 0.000 & 0.018 \\
$\Gamma =3.2$ & 1.000 & 0.024 & 0.001 & {\bf{0.041}} \\
$\Gamma =3.3$ & 1.000 & 0.035 & 0.002 & 0.058 \\
$\Gamma =3.4$ & 1.000 & {\bf{0.049}} & 0.003 & 0.080 \\
$\Gamma =3.5$ & 1.000 & 0.066 & 0.005 & 0.105 \\
$\Gamma =4$ & 1.000 & 0.195 & 0.031 & 0.266  \\
$\Gamma =4.1$ & 1.000 & 0.227 & {\bf{0.040}} & 0.296  \\
$\Gamma =4.2$ & 1.000 & 0.260 & 0.051 & 0.333 \\ \hline
\end{tabular}
\end{table}

Rather than a priori choose the subgroups, we might want to use the data to choose them.  However, we need to correct for the inherent multiplicity in having looked at the data to choose the subgroups.  We will discuss two procedures for doing so, splitting and using secondary information.

\subsubsection{Splitting: using a planning sample to choose which subgroups to analyze}
\label{split.sample.effect.modification}

As was done in Section \ref{splitting.obstetric} for choosing outcomes, one data-based strategy for choosing subgroups is splitting
\cite{cox1975note,heller2009split,lee2021discovering}.  Following the suggestion of \cite{lee2021discovering}, we split the sample randomly into a 25\% planning sample to choose the subgroups and then used the remaining 75\% analysis sample to test for a treatment effect in the chosen subgroups.  For the malaria study, to choose the subgroups using the planning sample, we computed the ranks of the matched pair differences between treatment and control in the malaria parasite outcome, regressed those ranks on gender and average age in the matched pair\footnote{Gender was exactly matched for all pairs.  Age was exactly matched in 94\% of pairs and differed by at most three years in any pair.} using a regression tree \cite{breiman1984classification,zhang2010recursive}, and chose the subgroups formed by the leaves of the tree.  
The regression tree contained one split at age $\geq 9.5$, so the two subgroups are age $<9.5$ and age $\geq 9.5$.  Using these two subgroups, we tested the null hypothesis of no treatment effect by combining the subgroup $p$-values on the analysis sample using the truncated product test (truncation point of 0.2).  The results are shown in the ``Splitting'' column in Table \ref{malaria.sensitivity}.  There is evidence for a treatment effect up to unmeasured confounding $\Gamma =4.1$.  Using the data to choose the subgroups with splitting reduced sensitivity compared to not using subgroups ($\Gamma = 1.9$) or using a priori chosen subgroups ($\Gamma =3.4$).

When the null hypothesis of no treatment effect at all is rejected, then we can test for no treatment effect in each subgroup, rejecting when the $p$-value is less than $\frac{0.05}{\# \mbox{subgroups} -1}$\cite{rosenbaum2008testing}.  There is evidence for an effect in both subgroups, age $<9.5$ and age $\geq 9.5$, for unmeasured confounding up to $\Gamma =1.7$ and evidence for an effect among age $<9.5$ for unmeasured confounding up to $\Gamma = 4.1$.

\subsubsection{Using secondary aspects of the data to choose which subgroups to analyze}
\label{effect.modification.secondary}

A second data-based strategy for choosing which subgroups to analyze is to use exploratory techniques on ``secondary'' aspects of {\it{all}} of the current data \cite{hsu2013effect,hsu2015strong,lee2018discovering}. The secondary aspects are chosen in such a way that an analysis that would control the familywise error rate if the groups were chosen a priori also controls the familywise error rate even though the groups were chosen using the data.  A simple version of this strategy, hereafter called the CART method, is to construct matched treated-control pairs based on the measured confounders and then fit a CART regression tree \cite{breiman1984classification} where the outcome for the tree is the absolute difference in outcomes in a matched pair between the treated and control unit (or the rank of the absolute difference) and the covariates are averages in a matched pair of variables that define the potential subgroups of interest, e.g., age and gender in the malaria study.  Note that the absolute differences in outcomes, a ``secondary'' aspect of the data, are used in building the CART tree rather than the ``primary'' aspect of the signed treated minus control differences.  However, this secondary aspect of the data potentially contains some information to find subgroups that are affected differently by the treatment.  Suppose the treatment minus control differences in outcome for the $i$th pair, denoted by $Y_i$, follow the model $Y_i=\rho ({\bf{x}}_i) + \epsilon_i$ where ${\bf{x}}$ are covariates and $\rho$ is a function such that $\rho ({\bf{x}})\geq 0$ for all ${\bf{x}}$ (i.e., the treatment effective is nonnegative) and $\epsilon$ are independent and identically from a mean zero continuous symmetric distribution (e.g., a standard normal distribution) \cite{hsu2013effect}.  We cannot estimate $\rho ({\bf{x}})$ from looking at the absolute treatment minus control differences, $|Y_i|$, but we do know that if $\rho ({\bf{x}}_i)>\rho ({\bf{x}}_{i'})$, then  $|Y_i|$ is stochastically larger than $|Y_{i'}|$  \cite{jogdeo1977association}.  This suggests that if the tree finds subgroups with larger $|Y_i|$, those subgroups might have larger $Y_i$, i.e., larger treatment effects.

The leaves of the CART tree become the subgroups.  The signs of the differences are then remembered in an analysis that treats the subgroups as if they were known a priori.  How can it be valid to treat the subgroups as if they were known a priori? Consider the simplest case of a paired randomized experiment \cite{lee2018discovering}.  If the null hypothesis of no effect of any kind were true, different random assignments would always yield the same absolute treated minus control outcome difference in matched pairs, only the signs would flip between matched pairs.  Hence, the CART tree that is built from looking at the absolute differences would not change between random assignments.  It is as if the subgroups were known a priori and the subgroups can be treated as known in testing hypotheses.  Things become more complicated when testing a null hypothesis about no effect in one subgroup (e.g., no effect among females)  but not among the whole population.  Then, even if this null hypothesis is true, different random assignments might alter the absolute treated minus control differences among pairs that are not in the subgroup considered in the hypothesis (e.g., among a pair of males) and thus the CART tree might be altered.  However, \cite{hsu2015strong} demonstrate two key facts:
\begin{enumerate}
\item[(1)] If the null hypothesis of no effect in a subgroup is true, the conditional distribution of a test statistic for that subgroup given that the CART tree contains that subgroup is the same as the randomization distribution over all random assignments.  In this sense, the instability of the tree over repeated randomizations has not distorted the conditional distribution of treatment
assignments in groups with no treatment effect.
\item[(2)] If a method is applied to test the subgroups chosen by the CART tree that would strongly control the familywise error rate at $\alpha$ with a priori fixed groups, then applying the method to the subgroups chosen by CART also controls the familywise error rate at $\alpha$.\footnote{\cite{karmakar2018false} shows that (2) also works for the false discovery rate {\textendash} if a method is applied that controls the false discovery rate if the groups were known a priori, then applying the method to the subgroups chosen by CART also controls the familywise error rate at $\alpha$.}
\end{enumerate}

\begin{figure}
\begin{center}
\includegraphics{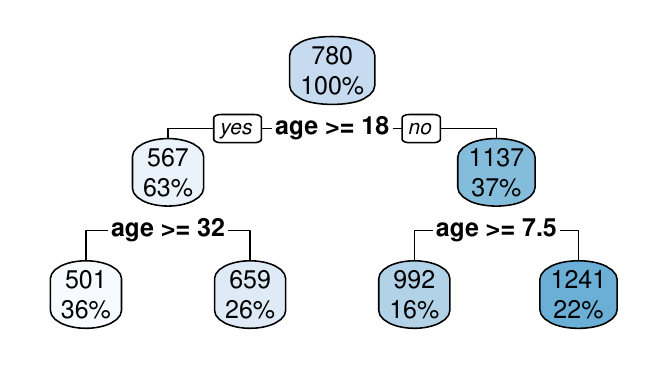}
\caption{For the malaria study, regression tree on age and gender using the rank of the absolute difference in a matched pair between the treated outcome and control outcome (difference in {\it{Plasmodium falciparum}} malaria parasites between the treatment and pre-treatment period).  The numbers in the circles are average rank of the absolute differences and the percentage of the sample in the node of the tree}
\label{absolute_tree}
\end{center}
\end{figure}

Figure \ref{absolute_tree} shows for the malaria data the fitted regression tree of the rank of absolute difference in the outcome between treatment and control on age and gender.  The tree suggested four subgroups: (i) age $<7.5$; (ii) age between 7.5 and 18; (iii) age between 18 and 32; (iv) age at least 32.  We tested the sharp null hypothesis of no treatment effect at all by using the truncated product test (truncation point of 0.2) with these four subgroups.  The null hypothesis is rejected for unmeasured confounding up to $\Gamma =3.2$ (last column of Table \ref{malaria.sensitivity}).

In the malaria study, the CART method  
was more sensitive to bias ($\Gamma =3.2$) than both an a priori choice of subgroups ($\Gamma =3.4$) and splitting ($\Gamma =4.1$).  In a simulation study, \cite{lee2018powerful} compared CART to the submax method that uses an a priori choice of subgroups.  They found CART was cautious about forming groups, so it failed to capitalize on moderate differential subgroup effects with a loss of power in some situations; however, that also meant that CART rarely paid a price for multiple testing when there was no difference in the effects between subgroups.

CART and other data adaptive methods for choosing subgroups like splitting might be particularly useful when there are many plausible subgroups as compared to methods that make an a priori choice of a few subgroups like the submax method.  For instance, in a study of the effect of having surgery at a hospital with superior nursing on surgical mortality \cite{lee2018discovering}, potential variables that could be used to form subgroups included 26 clusters of procedures (e.g., breast procedures, gall bladder procedures, stomach procedures), two comorbidities (congestive heart failure and chronic obstructive pulmonary disease), whether the patient was admitted to the hospital as an emergency admission and whether the patient was $>75$ years old.  Just among the 26 procedure clusters, there are many ways to group them; for instance, there are  $2^{26}-1=$ 67,108,863 ways
to split them into two groups.  CART built a tree with five groups. 
 For example, one group was a set of more serious procedures without an emergency admission among patients without congestive heart failure \cite{lee2018discovering}.  CART and the submax methods may be combined, e.g., an investigator may combine a few subgroups selected a priori with a few subgroups suggested by CART, applying the submax method to all of these subgroups.

\subsubsection{Discussion: it is important to notice subgroups with larger treatment effects in
observational studies}

In an observational study of treatment effects, there is invariably concern that an ostensible treatment effect is not actually an effect that is caused by the treatment, but rather some unmeasured bias distinguishing treated and control groups. Larger treatment effects are more insensitive to such concerns than smaller effects, i.e. larger biases measured by $\Gamma$ would need to be present to explain a larger treatment effect as spurious.  When the treatment effect is larger in some subgroups of the data than others, then by using methods which consider the possibility that there may be different treatment effects in different subgroups, we can often find stronger evidence to reject the null hypothesis of no treatment effect. For the malaria example, the combined test that does not consider subgroups was only insensitive to unmeasured confounding up to $\Gamma =1.9$ whereas three methods that consider subgroups {\textendash} an a priori choice of subgroups, splitting (using a planning sample to choose which subgroups to analyze) and using CART on secondary aspects of the data to choose which subgroups to analyze {\textendash} all were considerably more insensitive, up to $\Gamma = 3.4$, $4.1$ and $3.2$ respectively.  The methods that consider subgroups also provided information about in which subgroups there was evidence for a treatment effect that was less sensitive to unmeasured confounding, e.g., splitting suggested there was stronger evidence of a treatment among younger children (age $<9.5$) than among older children and adults (age $\geq 9.5$).  Such a discovery is important in three ways.  First, the finding about the affected subgroup is typically important in its own right as a description of that subgroup. Second, if there is no evidence of an effect in the complementary subgroup, then that may be news as well.  Third, if a sensitivity analysis convinces us that the treatment does indeed cause effects in one subgroup, then this demonstrates the treatment does sometimes cause effects, and it makes it somewhat more plausible that smaller and more sensitive effects in other subgroups (e.g., age $\geq 9.5$) are causal and not spurious. This is analogous to the situation in which we discover that heavy smoking causes lots of lung cancer, and are then more easily convinced that secondhand smoke causes some lung cancer, even though the latter effect is much smaller and more sensitive to unmeasured bias \cite{lee2018powerful}.

One caution in using data adaptive subgroup searching methods like CART and splitting 
that search for subgroups in which the treatment effect is less sensitive to bias from unmeasured confounding is that the bias may be greater in some subgroups than others.  If there was actually no treatment effect but the bias differed between subgroups, then these searching methods would report treatment effects that are less sensitive to bias than the combined test.  This is the opposite of what we want when there is no treatment effect.  When using searching methods, one should examine which subgroup(s) has been found to be least sensitive to bias and think about whether treatment assignment in that subgroup may be more biased than others.
In the malaria study, searching methods identified subgroups differing by age.  It is plausible that the bias does not differ between age subgroups because the treatment was applied to villages and so any bias in how the treated villages were chosen, e.g., political influence of the village chiefs, would apply equally to young and old (this assumes that the villages do not differ in their age distributions or that any differences in the age distribution are not correlated with an unmeasured confounder).

\begin{figure}
\begin{center}
\includegraphics{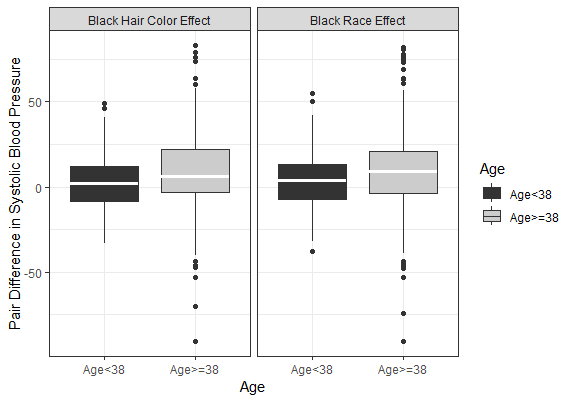}
\caption{For pairs matched on age and gender, boxplots of black hair vs. non-black hair differences in systolic blood pressure and black race vs. white race differences in systolic blood pressure.}
\label{race.bloodpressure}
\end{center}
\end{figure}

As an example of where the bias in subgroups might differ, consider the effect of black hair color on systolic blood pressure for the data described in Section \ref{selective_bias} where we pair matched people with black hair to people with non-black hair on age and gender.  Assuming no unmeasured confounding, there is strong evidence that black hair causes higher blood pressure {\textendash} the $p$-value is $<0.0001$ and the 95\% confidence interval is $(4.0,7.5)$ mmHg.  The effect is insensitive to bias up to $\Gamma =1.62$.  Using the CART method of Section \ref{effect.modification.secondary}, we found two subgroups, age less than $38$ and age $38$ or greater.  The left panel of Figure \ref{race.bloodpressure} shows the effect of black hair appears somewhat greater among those age $38$ or greater.  Using the truncated product test with the two subgroups, there is evidence of an effect of black hair up to $\Gamma =1.71$.  Presumably there is no actual effect of black hair on blood pressure and the apparent effect is due to the unmeasured confounder race {\textendash} African Americans have higher blood pressure than whites \cite{lackland2014racial}.  The use of age subgroups makes the results misleadingly more insensitive to bias because there is more of a racial difference in blood pressure for older people, see the right panel of Figure \ref{race.bloodpressure}.

The CART method and splitting look for subgroups in which the treatment effect might be different and then on those subgroups, carry out tests for treatment effects that control for multiple testing and the fact that the data was examined to find the subgroups.
A different approach to addressing treatment effect heterogeneity is to estimate conditional average treatment effects, conditional on a set of covariates.  For example, \cite{wager2018estimation} proposed the Causal Forest method to estimate covariate-specific treatment effects using random forests, \cite{su2009subgroup} used recursive partitioning to estimate treatment effects across subpopulations and \cite{hill2011bayesian} and \cite{hahn2020bayesian} developed Bayesian regression tree approaches.  These methods do not explicitly form subgroups but can be used to provide estimates of the treatment effect for any possible subgroup by averaging their estimates of the treatment effect for each subject in a subgroup over the subjects in the subgroup in the data set.  If subgroups of interest were specified in advance, then these methods could be combined with multiple testing methods to provide valid confirmatory inferences that control for multiple testing.  But if the subgroups of interest are not specified in advance, than these methods might be regarded as more exploratory than confirmatory.


For the CART method and splitting that use the data to form subgroups, one could ask, did we find the ``true groups''?  Arguably, there are no ``true groups'' but the division of subjects into groups is helpful for thinking about the strength of the evidence about a treatment effect and its practical implications.  However, if one's goal is to use a study to develop individualized treatment rules that say whether to treat a subject or not based on the subject's characteristics, as in precision medicine, then the ``true groups'' of interest are those for which treatment is beneficial and those for which it is not.  See \cite{kosorok2019precision} for a review of statistical methods for precision medicine.  For selecting and ranking individualized treatment rules from observational studies in which there might be unmeasured confounding, \cite{zhang2021selecting} develop sensitivity analysis methods.  For example, for the malaria study, consider six possible rules, $r_0, r_1,\ldots ,r_5$ where $r_i$ assigns treatment to the youngest $i\times 20\%$ of individuals.  Specifically, the minimum 20\%, 40\%, 60\%, 80\% and maximum of age are 0 (newborn), 7, 20, 31, 43 and 73 years old.  \cite{zhang2021selecting} find that if there were no unmeasured confounding, then a 95\% confidence set for the best individualized treatment rule is $\{ r_3,r_4,r_5\}$ but allowing for unmeasured confounding up to $\Gamma =2$ (an unmeasured confounder could double the odds of receiving treatment), a 95\% confidence set for the best individualized treatment rule contains four rules $\{ r_2, r_3,r_4,r_5\}$.

\section{Open Problems in Observational Study Protocols}
\label{open.problems}

In this section, we mention a few open problems related to observational study protocols (we also mentioned one open problem on assessing the  evidence for an elaborate theory at the end of Section \ref{elaborate.theories}).


\subsection{Splitting}

In Section \ref{sec: different.considerations}, we demonstrated that splitting may be a useful approach for selecting outcomes and subgroups in an observational study.  Some problems that have not been well explored for the approach include (i) what is a good fraction for the planning sample based on the sample size?; (ii) how should we decide how many and which outcomes should be carried forward from the planning sample to test in the analysis sample?; and (iii) instead of just having one planning sample and one analysis sample, how can we use sequential analysis methods 
to allow for multiple possible planning samples before deciding on an analysis sample?

\subsection{Selective Interpretation}
\label{selective.interpretation.open.problem}

In the mountaintop mining study, the primary analysis yielded insignificant results but an aspect of the secondary analysis yielded significant results and we thought of a post hoc explanation for why this pattern could be consistent with a true treatment effect.  How do we best summarize the conclusions of a study in this common situation?  In the discussion section of our mountaintop mining paper \cite{small2021surface}, we wrote
\begin{quote}
In our primary analysis of the period 1999–2011 and our secondary  analysis of the period 1990–1998, we did not find evidence of an effect of surface mining on low birthweight; however, in our secondary analysis  of the years 2012–2017, we did find evidence that surface mining was associated with low birthweight.
\end{quote}
but in the conclusion section, we wrote
\begin{quote}
We examined the hypothesis that surface mining activity in Central Appalachia contributes to low birth weight using an observational study. We found evidence in secondary analyses that surface mining was associated with increased odds of low birth weight in 2012–2017 and did not find such evidence for earlier time periods. A potential explanation for this pattern of association is that surface mining caused an increase in low birth weight but its onset was delayed.
\end{quote}
In this conclusion, did we pay too much heed to a secondary analysis finding?  Did we selectively interpret the findings?  It would be useful to have more empirical research on how readers interpret language in abstracts and conclusions of papers.

An extreme remedy for selective interpretation would be to, in the protocol, write the paper with blanks for the results and different discussions/conclusions that would be presented based on what the results are (like writing a {\it{Choose Your Own Adventure}} book \cite{packard1979cave}).  Potential problems with this approach include that it might be harder to think insightfully about results that hypothetically could occur vs. those that did occur and it is time consuming.  Nevertheless the approach may be worth trying.






\section{Summary}

Just as in a randomized trial, in an observational study, a protocol is an important part of making the study reliable.  We have discussed common features of good observational study and randomized trial protocols as well as differences.  We also mentioned some open methodological problems for observational study protocols.  Further research on methods for observational study protocols would be useful.

\bibliographystyle{imsart-number} 
\bibliography{bibtex_protocol_paper}       


\end{document}